\documentclass[aps,twocolumn,nofootinbib,preprintnumbers,eqsecnum,superscriptaddress]{revtex4}

\usepackage[dvips]{color}
\usepackage[normalem]{ulem}
\usepackage{amsmath}
\usepackage{enumerate}
\usepackage{amsfonts}
\usepackage{yfonts}
\usepackage{color}

\usepackage{graphicx}
\usepackage{bm}

\usepackage{epsfig}
\usepackage{slashed}

\def\({\left(}
\def\){\right)}
\def\[{\left[}
\def\]{\right]}
\def\<{\langle}
\def\>{\rangle}

\def\Tr{\mathop{\rm Tr}}

\newcommand\half{{\ensuremath{\frac{1}{2}}}}
\newcommand\p{\ensuremath{\partial}}

\newcommand\vev[1]{{\ensuremath{\left\langle{#1}\right\rangle}}}

\newcommand{\be}{\begin{equation}}
\newcommand{\ee}{\end{equation}}
\newcommand{\bea}{\begin{eqnarray}}
\newcommand{\eea}{\end{eqnarray}}
\newcommand{\bwt}{\begin{widetext}}
\newcommand{\ewt}{\end{widetext}}

\newcommand{\bi}{\begin{itemize}}
\newcommand{\ei}{\end{itemize}}
\newcommand{\ben}{\begin{enumerate}}
\newcommand{\een}{\end{enumerate}}
\newcommand{\bca}{\begin{cases}}
\newcommand{\eca}{\end{cases}}
\newcommand{\bln}{\begin{align}}
\newcommand{\eln}{\end{align}}
\newcommand{\bst}{\begin{split}}
\newcommand{\est}{\end{split}}

\newcommand\Sig{\Sigma}
\newcommand\lam{\lambda}

\newcommand\da{{\dagger}}

\def\th{{\theta}}

\newcommand\ov{\over}
\newcommand\ha{{\half}}

\def\le{\left}
\def\ri{\right}

\newcommand\sC{{\ensuremath{{\mathcal C}}}}

\newcommand\sN{{\ensuremath{{\mathcal N}}}}

\newcommand\sR{{\ensuremath{{\mathcal R}}}}

\newcommand\sW{{\mathcal W}}

\renewcommand{\Im}{\textrm{Im}\,}
\renewcommand{\Re}{\textrm{Re}\,}

\begin{document}

\title {Transverse Momentum Broadening and the Jet Quenching Parameter, Redux}

\preprint{MIT-CTP-4153}

\author{Francesco D'Eramo, Hong Liu and Krishna Rajagopal}

\affiliation{Center for Theoretical Physics, \\
Massachusetts
Institute of Technology, \\
Cambridge, MA 02139
}

\bigskip
\bigskip

\vspace{5cm}

\begin{abstract}

We use Soft Collinear Effective Theory (SCET) to analyze the transverse momentum broadening, or diffusion in transverse momentum space, of an energetic parton propagating through quark-gluon plasma.   
Since we neglect the radiation of gluons from the energetic parton, we can only discuss momentum broadening, not parton energy loss.  The interaction responsible for momentum broadening in the absence of radiation is that between the energetic (collinear) parton and the Glauber modes of the gluon fields in the medium.  We derive the effective Lagrangian for this interaction, and we show that the probability for picking up transverse momentum $k_\perp$ is given by the Fourier transform of the expectation value of two transversely separated light-like path-ordered Wilson lines.  This yields a field theoretical definition of the jet quenching parameter $\hat q$, and shows that this can be interpreted as a diffusion constant.     We close by revisiting the calculation of $\hat q$ for the strongly coupled plasma of ${\cal N}=4$ SYM theory, showing that previous calculations need some modifications that make them more straightforward and do not change the result.

\end{abstract}

\date{June 7, 2010}

\maketitle

\section{Introduction}

One of the central discoveries made in experimental heavy ion collisions at the Relativistic Heavy Ion Collider (RHIC) at Brookhaven National Laboratory is that the droplets of quark-gluon plasma produced in these collisions are sufficiently strongly coupled that they are 
able to ``quench jets''~\cite{Arsene:2003yk,Adler:2003ii,Back:2003ns,Adams:2003im}.  That is, when a very energetic quark or gluon (energetic enough that if it were in vacuum it would manifest itself as a jet, and with an energy that is much greater than the temperature of the medium in which it finds itself) plows through the strongly coupled plasma, it loses sufficient energy that few high momentum hadrons are seen in the final state.  This phenomenon manifests itself in many observables, including a depletion of the overall number of high transverse momentum hadrons and the absence of high transverse momentum hadrons back-to-back with a single high momentum hadron that the experimentalists have triggered on. The suite of observables collectively referred to as jet quenching makes it possible to study how parton fragmentation is affected by the presence of a strongly coupled plasma and of how the medium responds to the energy and momentum that a fragmenting parton dumps into it.
After a very energetic parton is produced, unless it is produced at the edge of the fireball heading outwards it must propagate through as much as 5-10 fm of the hot and dense medium produced in the collision.  Because the production cross-sections for hard partons are well determined (both by perturbative QCD calculations and by experimental measurements in proton-nucleus collisions) these hard partons serve as well-calibrated probes of  the strongly coupled plasma whose properties we are interested in.   At present at RHIC, the highest energy partons that are available for use as probes of the medium have transverse momenta of several tens of GeV.  When heavy ion collision experiments begin at the Large Hadron Collider at CERN in Geneva, experimentalists there will be able to study the interaction of partons with a few hundred GeV momenta with the medium produced in those collisions.

The presence of the strongly coupled medium results in the hard parton losing energy and changing the direction of its momentum.  The change in the direction of its momentum is often referred to as ``transverse momentum broadening'', a phrase that needs explanation.  ``Transverse'' here and throughout the remainder of this paper means perpendicular to the original direction of the hard parton.  (In the preceding paragraph, we used ``transverse'' in its more standard sense, meaning perpendicular to the beam direction.)  ``Broadening'' refers to the effect on a jet when the directions of the momenta of many hard partons within it are kicked;  averaged over many partons in one jet, or perhaps in an ensemble of jets, there is no change in the mean momentum but the spread of the momenta of the individual  partons broadens.

In the high parton energy limit, the parton loses energy dominantly by 
inelastic processes that are the QCD analogue of bremsstrahlung: the parton radiates gluons as it interacts with the medium.  It is a familiar fact from electromagnetism that bremsstrahlung dominates the loss of energy of an electron moving through matter in the high energy limit.
The same is true in calculations of QCD parton energy loss in the high-energy 
limit, as established first in Refs.~\cite{Gyulassy:1993hr,Baier:1996sk,Zakharov:1997uu}.
The hard parton undergoes multiple inelastic interactions with the spatially extended medium, and this induces gluon bremsstrahlung.     It is crucial to the calculation of this radiative energy loss process that the incident hard parton, the outgoing parton, and the radiated gluons are all continually being jostled by the medium in which they find themselves:  they are {\it all} subject to transverse momentum broadening.    

The transverse momentum broadening of a hard parton is described by $P(k_\perp)$, defined as the probability that after propagating through the medium for a distance $L$ the hard parton has acquired 
transverse momentum $k_\perp$.\footnote{Throughout this paper, $k_\perp$ is the two-dimensional vector $\vec{k}_\perp$ in transverse momentum space; we shall drop the vector symbol for notational convenience.  When we write $d^2k_\perp$ we mean $dk_x dk_y$.  And, $k_\perp^2$ will mean $\vec{k}_\perp\cdot\vec{k}_\perp$. We will also represent two-dimensional vectors in transverse coordinate space by $x_\perp$ and $y_\perp$, again without the vector symbols.} $P(k_\perp)$ depends on $L$, but we shall not make this dependence explicit in the notation.
For later convenience, we shall choose to normalize $P(k_\perp)$ as follows:
\begin{equation}
 \int \frac{d^2 k_{\perp}}{(2\pi)^2} P (k_{\perp}) = 1\ .
\label{eq:Pnormalization}
\end{equation}
{}From the probability density $P(k_\perp)$, it is straightforward to obtain the mean transverse momentum picked up by the hard parton per unit distance travelled (or, equivalently in the high parton energy limit, per unit time):
\begin{equation}
\hat q \equiv \frac{\langle k_\perp^2 \rangle}{L} =
\frac{1}{L} \int \frac{d^2 k_{\perp}}{(2\pi)^2} k_\perp^2 P (k_{\perp})\ .
\label{qhatFirstTime}
\end{equation}
The quantity $\hat q$ is called the ``jet quenching parameter'' because it plays a central role in calculations of radiative parton energy loss~\cite{Baier:1996sk,Zakharov:1997uu,Wiedemann:2000za,Gyulassy:2000er,Guo:2000nz,Wang:2001ifa,Arnold:2002ja}, reviewed in Refs.~\cite{Baier:2000mf,Kovner:2003zj,Gyulassy:2003mc,Jacobs:2004qv,CasalderreySolana:2007zz,Accardi:2009qv,Wiedemann:2009sh,Majumder:2010qh}.  Consequently, $\hat q$ is  thought of as a (or even the) property of the strongly coupled medium that is ``measured'' (perhaps constrained is a better phrase) by radiative parton energy loss and hence jet quenching. But, it is important to note that $\hat q$ is {\it defined} via transverse momentum broadening only.  Radiation and energy loss do not arise in its definition, although they are central to its importance.

The calculation of parton energy loss and transverse momentum broadening involves widely separated scales.  The energy of the hard parton arises in both, as does the soft scale $T$ characteristic of the medium.  In the case of radiative parton energy loss, the momentum of the radiated gluon transverse to that of the incident parton represents a third, intermediate, scale.  We can ultimately hope for a factorized description, with physics at each of these scales cleanly separated at lowest nontrivial order in a combined expansion in the small ratio between these scales and in the QCD coupling $\alpha$ evaluated at scales which become large in the high parton energy limit.  And, most importantly, we can aspire to having a formalism in which
corrections to this factorization are calculable systematically, order by order in these expansions.  No current theoretical formulation of jet quenching calculations is manifestly systematically improvable in this sense.  In this paper we take a small step toward such a description: we formulate the calculation of transverse momentum 
broadening and the jet quenching parameter in the language of Soft Collinear Effective Theory (SCET)~\cite{Bauer:2000ew} which has rendered the calculation of many other processes involving soft and collinear degrees of freedom systematically improvable.  
We are following in the footsteps 
of Idilbi and Majumder~\cite{Idilbi:2008vm}, who made the first attempt to extend SCET to describe hard jets in a dense medium and in so doing realized that transverse momentum broadening is induced by the interaction between the hard parton and gluons from the medium whose lightcone momenta are much softer than $T$.  Gluons in this kinematic regime, which we shall define precisely in Section II, are conventionally called ``Glauber gluons''.  The analysis of Ref.~\cite{Idilbi:2008vm} builds upon the earlier analysis of transverse momentum broadening in Ref.~\cite{Majumder:2007hx}.

In Section II, we use the formalism of SCET to set up the problem of transverse momentum broadening without radiation. We reproduce the result of Ref.~\cite{Idilbi:2008vm} that Glauber gluons are responsible.  Extending the calculation to include radiation, and hence parton energy loss, is an obvious next step, but we leave it to future work.  (Note that neglecting radiation is not justified by any controlled approximation.) Sections III, IV and V contain our calculation of $P(k_\perp)$.  In Section III we use the optical theorem to relate $P(k_\perp)$ to suitable forward scattering amplitudes. In Section IV we use SCET to set up the effective Lagrangian and Feynman rules needed to evaluate the relevant amplitudes, and in Section V we complete the computation.  The result we obtain agrees with an expression first obtained by Casalderrey-Solana and Salgado and by Liang, Wang and Zhou 
using different methods~\cite{CasalderreySolana:2007zz,Liang:2008vz}.  
The bigger payoff from a SCET calculation will come once radiation is included and once the analysis is pushed beyond the present leading order calculation.

With $P(k_\perp)$ in hand, it is easy to obtain the jet quenching parameter $\hat q$, as we describe in Section VI.  We find that $P(k_\perp)$, and hence $\hat q$, are determined by the thermal expectation value of the trace of the product of two light-like Wilson lines separated by a distance $x_\perp$ in the perpendicular direction, which we denote $\sW(x_\perp)$.  $P(k_\perp)$ is in fact the Fourier transform of this quantity.  Note that $\sW(x_\perp)$ depends only on the properties of the strongly coupled medium.  It is independent of the energy of the hard parton, meaning that so are $P(k_\perp)$ and $\hat q$ in the limit in which this hard parton energy is taken to infinity.  
Transverse momentum broadening without radiation thus does ``measure'' a field-theoretically well-defined property of the strongly coupled medium.  This is the kind of factorization that we hope to find in a systematically improvable calculation once radiation is included.

Crucially, and as a consequence of our field-theoretical formulation, we find that the ordering of operators in the expectation value $\sW(x_\perp)$ is {\it not} that of a standard Wilson loop --- the operators (like the color matrices)  are path ordered, whereas in a standard Wilson loop operators are time ordered. This subtlety, which had not been noticed previously in the present context although the same ordering does arise in a different kinematic regime~\cite{CasalderreySolana:2007qw}, throws into question the calculation of $\hat q$ in the strongly coupled plasma of large-$N_c$ ${\cal N}=4$ supersymmetric Yang-Mills (SYM) theory reported in  Refs.~\cite{Liu:2006ug,Liu:2006he}.  In Section VII we therefore use gauge/gravity duality to repeat this calculation. The subtleties introduced in the gravity calculation, corresponding to the subtlety of operator ordering in the field theory, turn out not to change the results of Ref.~\cite{Liu:2006ug} for $\sW(x_\perp)$ and $\hat q$ in ${\cal N}=4$ SYM theory.  In fact, they solidify these results since it now turns out that there is  only one extremized string world sheet that is bounded by the light-like Wilson lines, and it is the one identified on physical grounds in Refs.~\cite{Liu:2006ug,Liu:2006he}.

In Section VIII we look ahead to future work.


\section{Set-up and Kinematics}

Consider an on-shell high energy parton with initial four momentum\footnote{The light cone coordinates are defined by $q^{\pm} = {1 \ov \sqrt{2}} (q^0 \pm q^3)$.
}
\begin{equation}
q_0 \equiv (q_0^+, q_0^-, q_{0 \perp}) = (0, Q, 0)
\end{equation}
propagating through some form of QCD matter, which for definiteness we will take to be quark-gluon plasma (QGP) in equilibrium at temperature $T$ although the discussion of this paper would also apply to propagation through other forms of matter. We will assume throughout that $Q$ is very much larger than the highest momentum scales that characterize the medium, which for the case of QGP means 
$Q\gg T$.  Thus, we have a small dimensionless ratio in our problem:
\begin{equation} \label{hilim}
\lam \equiv {T \ov Q} \ll 1 \ .
\end{equation}
Although as we have already noted we will use much of the SCET formalism developed 
by Idilbi and Majumder~\cite{Idilbi:2008vm}, there are important differences between our use of this formalism and theirs.
First, we shall analyze the propagation of the high energy parton in the frame in which the medium  through which it is propagating is at rest whereas they work in the Breit frame, as appropriate for their analysis of deep inelastic scattering on large nuclei.  
Second, our definition (\ref{hilim}) differs from the choice of $\lambda$ made in 
Ref.~\cite{Idilbi:2008vm}.  In their case, they choose $\lambda$ such that $\lambda^2 Q$ is somewhat greater than the characteristic momentum scale of the partons in the nuclear medium that they analyze whereas we have defined $\lambda$ such that $\lambda Q = T$.   

Our goal is to characterize the transverse momentum broadening of the hard parton by computing 
$P (k_\perp)$, the probability density for the transverse momentum $k_\perp$ acquired by the hard parton after it has propagated through the medium for a distance $L$.
In the high energy limit~\eqref{hilim}, there are three distinct contributions to the transverse momentum broadening of the hard parton:

\ben

\item 

Generic gluons in the medium, which we shall call soft gluons, have momenta
\begin{equation}
p_s \sim (T,T,T) \sim  Q (\lambda,\lambda,\lambda) \ .
\label{psoft}
\end{equation}
After the hard parton absorbs (or emits) a soft gluon from (into) the medium, its momentum becomes $\sim Q(\lambda,1,\lambda)$.  This means that it has picked up transverse momentum of order $\lambda Q\sim T$. In addition, however, the hard parton has been kicked off-shell by of order $\lambda Q^2$, meaning that this process is suppressed by a coupling  
$\alpha_s(\sqrt{T Q})$, which is small in the high energy limit (and, plausibly, for jets with $Q>100$~GeV in nucleus-nucleus collisions at the LHC.)
Subsequently, the off-shell parton radiates a gluon or gluons, possibly  with 
momenta $\sim Q(\lambda,1,\sqrt{\lambda})$.
A future analysis of these processes (which are suppressed by powers of a 
perturbative $\alpha_s$ but may nevertheless make an important contribution to radiative parton energy loss because the medium is dominantly soft gluons)
will require adding modes with momenta $\sim Q(\lambda,1,\lambda)$, and possibly $\sim Q(\lambda,1,\sqrt{\lambda})$, to the effective theory, namely modes that are not soft  and are also not the collinear or Glauber modes to which we now turn.

\item

The hard parton can absorb (or emit) gluons from (into) the medium that are parametrically softer than (\ref{psoft}). This induces momentum broadening without radiation.  
Specifically, consider gluons from the medium with momenta
 \begin{equation} \label{doM}
p \sim Q (\lambda^2, \lambda^2, \lambda)\ ,
\end{equation}
which are normally called ``Glauber gluons''.
After absorbing or emitting Glauber gluons, the momentum of the hard parton 
is of order  
\begin{equation}\label{CollinearDefn}
q \sim Q (\lambda^2, 1, \lambda)\ .
\end{equation}
We shall refer to modes with momenta of this parametric order as ``collinear''.
As above, the momentum broadening is of order $\lambda Q\sim T$.  But, here the parton is only off-shell by of order $\lambda^2 Q^2\sim T^2$. Further absorption or emission of Glauber gluons keeps the parton off-shell by of the same order.\footnote{As does absorption or emission of modes with momenta $\sim Q(\lambda^2,\lambda^2,\lambda^2)$, which are conventionally referred to as ``ultrasoft.''  Since in the present context they introduce no distinct physical effects, they are simply a subset of the Glauber modes and thus are included in our analysis.  So too are all modes whose momenta are proportional to even higher powers of $\lambda$.  Interestingly, modes with momenta $\sim Q(\lambda^2,\lambda,\lambda)$ also introduce no distinct physical effects.  They are not conventionally called Glauber gluons, but our analysis of Glauber gluons applies to them also.}
And yet, repeated absorption and emission of Glauber gluons continually kicks the hard parton and can result in significant transverse momentum broadening.
The interaction vertex of each Glauber gluon with the parton is governed by $\alpha_s(T)$ and so can be strongly coupled.
At a heuristic level, one can imagine Glauber gluons as a gluon background
surrounding the parton and as a result of frequent small kicks from this background, the parton will undergo Brownian motion in momentum space.  We shall see that in the high energy limit the trajectory of the parton in position space is not affected.


\item

The initial hard parton, with a collinear momentum $q$ as in (\ref{CollinearDefn}), can fragment into two collinear partons each with momenta of this order.  Since all three partons have virtuality of order $\lambda^2 Q^2\sim T^2$, this radiative process is not suppressed by any perturbatively small $\alpha_s$.  (Although fragmentation can occur in vacuum, in the medium the rate for  this process will be modified, as all three partons are continually interacting with the Glauber gluons from the medium.)  This radiative process is the dominant contribution to parton energy loss. It also contributes to transverse momentum broadening, but we defer the calculation of this contribution to the future, focussing in this paper on momentum broadening in the absence of radiation, as in 2.  We also defer consideration of the case where the initial parton itself is far off-shell, and consequently radiates.

\een

Let us recapitulate. Processes 1, 2 and 3 all yield transverse momentum broadening.  
We neglect process 3 in this paper, deferring the case where the number of collinear particles increases to a future analysis of radiative parton energy loss.  
Process 1
is suppressed by $\alpha_s$ evaluated at a high momentum scale meaning that in the $Q\rightarrow\infty$ limit process 2 makes a larger contribution to momentum broadening.  With this justification, we shall neglect process 1 throughout this paper.  But, process 2 is triggered by the Glauber gluons in the medium, and these are less numerous than the soft gluons that are the generic modes in the plasma.  So, even though process 2 dominates process 1
in the strict $Q\rightarrow\infty$ limit, process 1 may be relevant at the values of the hard parton momentum $Q$ that are accessible at RHIC and the LHC.  Clearly, before data can be confronted all 3 processes must be included.  We focus on non-radiative momentum broadening in the $Q\rightarrow\infty$ limit in this paper both because it is the easiest case to handle and because it provides the natural context in which the jet quenching parameter arises.

We shall be interested in a hard parton with initial momentum $Q(0,1,0)$ that interacts repeatedly with Glauber gluons; after the first such interaction the hard parton momentum is of order $Q(\lambda^2,1,\lambda)$, namely collinear.  

Before beginning our analysis, we should explain why we neglect gluons from the medium whose momenta scale with $\lambda$ in some way other than the Glauber gluons (that we include) or the soft gluons (whose neglect we have explained above).  Gluons in the medium for which some momentum component scales like a positive power of $Q$ are in principle present but they are suppressed by an exponentially small Boltzmann factor in the
$\lambda\to 0$ high energy limit, and can safely be neglected.  
The only other case to consider is gluons in the medium with momenta $\sim Q(\lambda,\lambda^2,\lambda)$.   
Interaction with these gluons from the medium increases the virtuality of the hard parton in the same way that interaction with the soft gluons does,  
and so should be 
treated in an analysis beyond that in this paper which includes radiative processes.

We shall use the language of Soft Collinear Effective Theory (SCET)~\cite{Bauer:2000ew} to set up this problem, since in the $\lambda\rightarrow 0$ limit we have a natural separation of scales, and a natural organization of the modes into kinematic regimes: collinear, soft, and Glauber.  The problem that we analyze in this paper really does not use the full power of SCET, however, since we shall only work to lowest nontrivial order in $\lambda$ and in $\alpha_s$ evaluated at any scale higher than $T$. Including radiative processes and the interaction with the soft gluons that is suppressed by $\alpha_s(\sqrt{TQ})$ 
would exercise more of the machinery of SCET than our analysis of transverse
momentum broadening in the absence of radiation will.

In our analysis we will essentially consider the Glauber gluon insertions as external background fields and average them over the thermal ensemble only at the end of the calculation.  That is, we first consider the propagation of the hard parton in the presence of one background field configuration, analyzing this problem including arbitrarily many interactions with the background field.  The nonperturbative physics of the medium does not enter this calculation.   We shall then stop, leaving unevaluated in our answer the quantity that arises that does depend on the physics of the medium, namely the jet quenching parameter
$\hat q$. As we shall show, $\hat q$ is determined by an average of certain light-like Wilson lines over background field configurations drawn from a thermal ensemble.  
  If the medium is weakly coupled, this average should be perturbatively calculable in thermal field theory.
If the medium is strongly coupled, the only nonperturbative method we know of evaluating this thermal average is gauge/gravity duality. In Section VII we revisit the computation of $\hat q$ in strongly coupled ${\cal N}=4$ Supersymmetric Yang-Mills theory.

\section{Optical theorem and transverse momentum broadening}

\begin{figure*}
 \begin{center}
\includegraphics[scale=0.4]{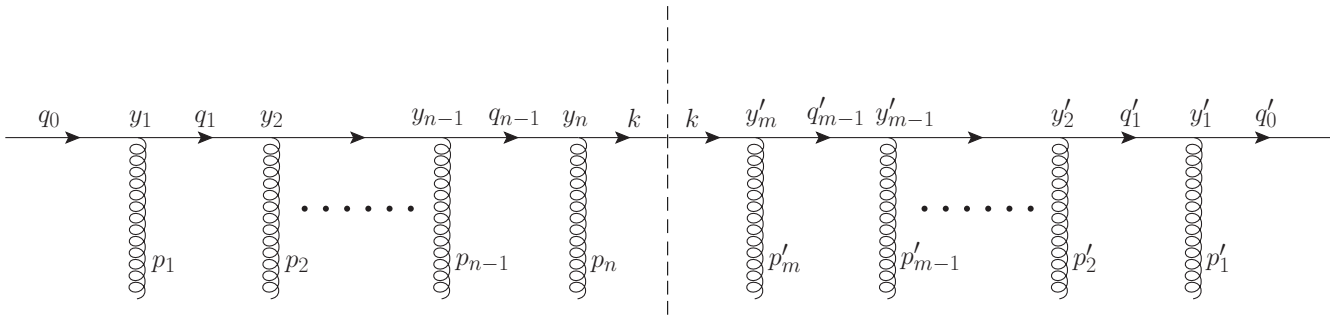}
\end{center}
\caption{Leading diagram for the transverse momentum broadening of a hard (collinear) parton, in the absence of any radiation. All the gluons are Glauber gluons.  We shall assume that all the Glauber gluons are drawn from an equilibrium thermal ensemble, neglecting the possibility that earlier interactions of the hard parton with the medium may change the medium in some way that affects the hard parton through later interactions.  That is, we are neglecting the possibility that the way that the medium recoils could influence the distribution of gluons seen by the hard parton.}
\label{fig:forw}
\end{figure*}

Our goal is to compute the probability distribution $P (k_{\perp})$, normalized as in (\ref{eq:Pnormalization}), which is the probability for the hard parton to have acquired
transverse momentum $k_{\perp}$ after its propagation through the medium for a distance $L$. 
The field theory tools we use to perform this computation are the $S$ matrix and the optical theorem. In particular, we cut the forward scattering amplitude to get the total probability for any interaction to have occurred.  If we add to this the probability that the hard parton propagates a distance $L$ through the medium without any interaction, we must obtain 1.

We shall imagine a cubic box with sides of length $L$ that is filled with the medium, and that satisfies periodic boundary conditions.
This leads to the quantization of momenta
\begin{equation}
 \mathbf{p} = \frac{2\pi}{L} \left(n_1,n_2,n_3\right)\ ,
\label{boxn}
\end{equation}
for integers $n_i$.
We consider a single-particle in state describing the incident hard parton. 
Because we are ignoring radiation, the only out states that we need consider are also single-particle states, which we shall label by the
quantized three-momentum $\mathbf{p}$ and a discrete set of quantum numbers $\sigma$ (mass, spin, charge, etc.). We use a Greek letter to denote the whole collection of $\mathbf{p}$ and $\sigma$, and we normalize the states as follows
\begin{equation}
\langle \alpha^\prime | \alpha \rangle = \delta_{\alpha^\prime \alpha}\ ,
\label{BoxNormalization}
\end{equation}
where $\delta_{\alpha^\prime \alpha}$ is a Kronecker delta. 

{}From now on we denote the single particle initial and final states by $|\alpha \rangle$ and $|\beta \rangle$, respectively. The $S$-matrix element $S_{\beta\alpha}$ is defined as the probability amplitude for the process $\alpha \rightarrow \beta$. Conservation of probability implies unitarity for the $S$-matrix:
\begin{equation}
 \sum_{\beta} \left|S_{\beta\alpha}\right|^2 = 1 \ .
\label{eq:Sunit}
\end{equation}
As usual, we first isolate the identity part of the S-matrix
\begin{equation}
S_{\beta \alpha} = \delta_{\beta\alpha} + i M_{\beta\alpha}\ ,
\label{eq:SdecApp}
\end{equation}
in so doing defining the interaction matrix element $M_{\beta\alpha}$.
{}From (\ref{eq:SdecApp}) we obtain
\begin{equation}
 \left|S_{\beta\alpha}\right|^2 = 
 \left\{ \begin{array}{lc}
|M_{\beta\alpha}|^2 & \quad\beta \neq \alpha \\
1- 2 \Im M_{\alpha\alpha} + |M_{\alpha\alpha}|^2 &\quad \beta=\alpha \ .
\end{array} \right.
\label{eq:SSquared}
\end{equation}
The unitarity condition (\ref{eq:Sunit}) then reads
\begin{equation}
2 \,\Im\, M_{\alpha\alpha} = \sum_{\beta} |M_{\beta\alpha}|^2\ .
\label{eq:imp1}
\end{equation}
At this formal level, (\ref{eq:imp1}) would still be valid if we were including the effects of radiation, meaning that final states $|\beta\rangle$ would include many particle states.

In the present paper, we are interested in computing the probability for the process $\alpha \rightarrow \beta$, where both states describe single particles with $v=1$ since we are taking $Q\rightarrow\infty$.
The final state $\beta$ differs from the initial state $\alpha$ only in its value of $k_{\perp}$. No other quantum numbers change. In particular, $\beta=\alpha$ corresponds to $k_{\perp}=0$. Thus, we have
\begin{equation}
\sum_{\beta} = L^2 \int \frac{d^2 k_{\perp}}{(2\pi)^2}\ ,
\label{eq:sumbeta}
\end{equation}
where the $L^2/(2\pi)^2$ comes from the box normalization,  in which summing over $\beta$ means summing over the values of the $n_1$ and $n_2$ in (\ref{boxn}).
And, in our context the
unitarity condition (\ref{eq:Sunit}) must be equivalent to the statement that $P(k_\perp)$ is normalized as in (\ref{eq:Pnormalization}).  Upon using (\ref{eq:SSquared}) and (\ref{eq:sumbeta}) to write (\ref{eq:Sunit}), comparison to (\ref{eq:Pnormalization}) therefore allows us  to identify
\begin{equation}
P(k_{\perp}) = L^2 \left\{ \begin{array}{lc}
|M_{\beta\alpha}|^2 &\quad \beta \neq \alpha \\
1- 2\, \Im M_{\alpha\alpha} + |M_{\alpha\alpha}|^2 &\quad \beta=\alpha \ .
\end{array} \right.
\label{eq:PvsS}
\end{equation}
Our strategy will be to first compute twice the imaginary part of the forward scattering amplitude, $2\,\Im M_{\alpha\alpha}$, by cutting the appropriate diagrams. Once we know $2\, \Im M_{\alpha\alpha}$ we can use the unitarity relation (\ref{eq:imp1}) to read off $\sum_\beta |M_{\beta\alpha}|^2$. Knowing (\ref{eq:sumbeta}), we will immediately be able to identify $P(k_\perp)$ for $k_\perp\neq 0$ (i.e. $\alpha\neq \beta$), and the normalization condition (\ref{eq:Pnormalization}) will then fix $P(0)$.

\begin{figure*}[t]
 \begin{center}
\includegraphics[scale=0.4]{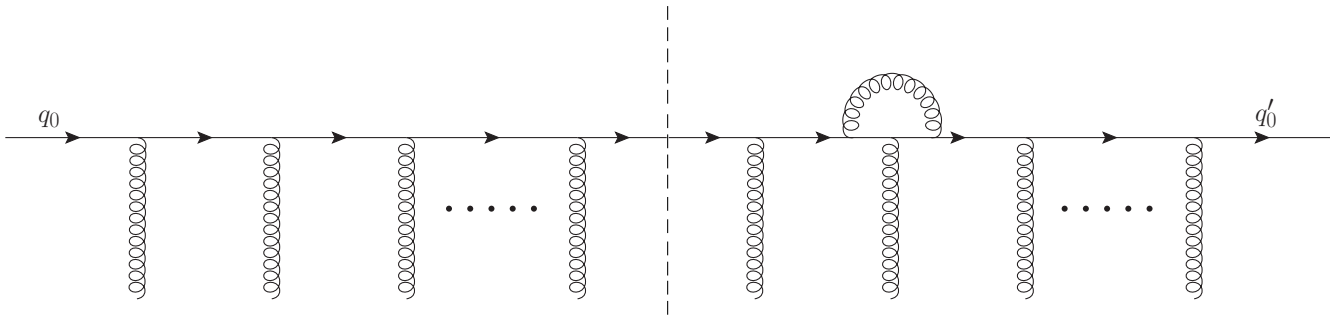}
\end{center}
\caption{Loop diagrams like this one vanish for kinematic reasons if the gluon in the loop is soft or Glauber.  If the gluon in the loop is collinear, cutting this diagram describes radiative processes with two collinear particles in the final state which we do not analyze in this paper.
}
\label{fig:loop}
\end{figure*}

The leading diagrams for the forward scattering amplitude are given in Fig.~\ref{fig:forw}, with the dashed line indicating where we cut the diagram and with all gluon lines being Glauber gluons.  Note that we include arbitrarily many Glauber gluon insertions.   In Fig.~\ref{fig:forw}, the Glauber gluons can be connected to the parton propagators in $(n+m)!$ ways, but evaluating the diagram yields an identical factor in the denominator, meaning that we can consider just one diagram for a given $m,n$.  The full amplitude is obtained by summing over all $m$ and $n$. If we denote twice the imaginary part of the amplitude depicted in Fig.~\ref{fig:forw} by $\mathcal{A}_{mn}$, the imaginary part of the forward scattering amplitude is then given by
 \be \label{eq:amp1}
 2\,\Im M_{\alpha\alpha}= \sum_{m=1,n=1}^{\infty} \mathcal{A}_{mn} \ .
 \ee
To compute $P(k_\perp)$ we will need to separate the integration over the transverse momentum $k_{\perp}$ at the cutting line. It will therefore be convenient to introduce $d^2 \mathcal{A}_{mn}/d^2 k_\perp$ via
\begin{equation}
\mathcal{A}_{mn} = \int \frac{d^2 k_{\perp}}{(2\pi)^2}\,\frac{d^2 \mathcal{A}_{m n}}{d^2 k_{\perp}} \ .
\end{equation}

We shall not consider loop diagrams  as in Fig.~\ref{fig:loop}. When the gluon in the loop is soft or Glauber (or in fact ultrasoft) the diagram is trivially zero in the Feynman gauge.  If we define the light cone vectors $\bar n^\mu$ (in the direction of motion of the hard parton) and $ n^\mu$ (useful in the next section) by
\be
\bar{n}\equiv \frac{1}{\sqrt{2}}\left(1,0,0,-1\right) \quad {\rm and}\quad n\equiv\frac{1}{\sqrt{2}}\left(1,0,0,1\right),
\label{nDef}
\ee
in Feynman gauge the vertices in the loop in Fig.~\ref{fig:loop} are given by
$\bar{n}^{\nu} \bar{n}^{\mu} g_{\mu\nu} = 0$. When the gluon in the loop is collinear the diagram is nonvanishing but we defer its evaluation to a future evaluation of radiative processes, since cutting this diagram across the loop describes a radiative process with two collinear particles in the final state, as in what we called process 3 in Section II.

\section{Effective Lagrangian and Feynman rules}

In this Section we describe the Feynman rules needed to compute the amplitude in 
Fig.~\ref{fig:forw}. For this purpose, we need the effective Lagrangian governing the interaction between collinear partons and  Glauber gluons~\cite{Idilbi:2008vm}. We shall sketch its derivation using SCET~\cite{Bauer:2000ew}.

The hard parton propagating through the medium can be a quark or gluon.  We shall first take it to be a quark, describing the derivation of the effective Lagrangian in full for this case.
We will then quote the result for the case where the hard parton is a gluon.

\begin{figure}[t]
 \begin{center}
\includegraphics[scale=0.5]{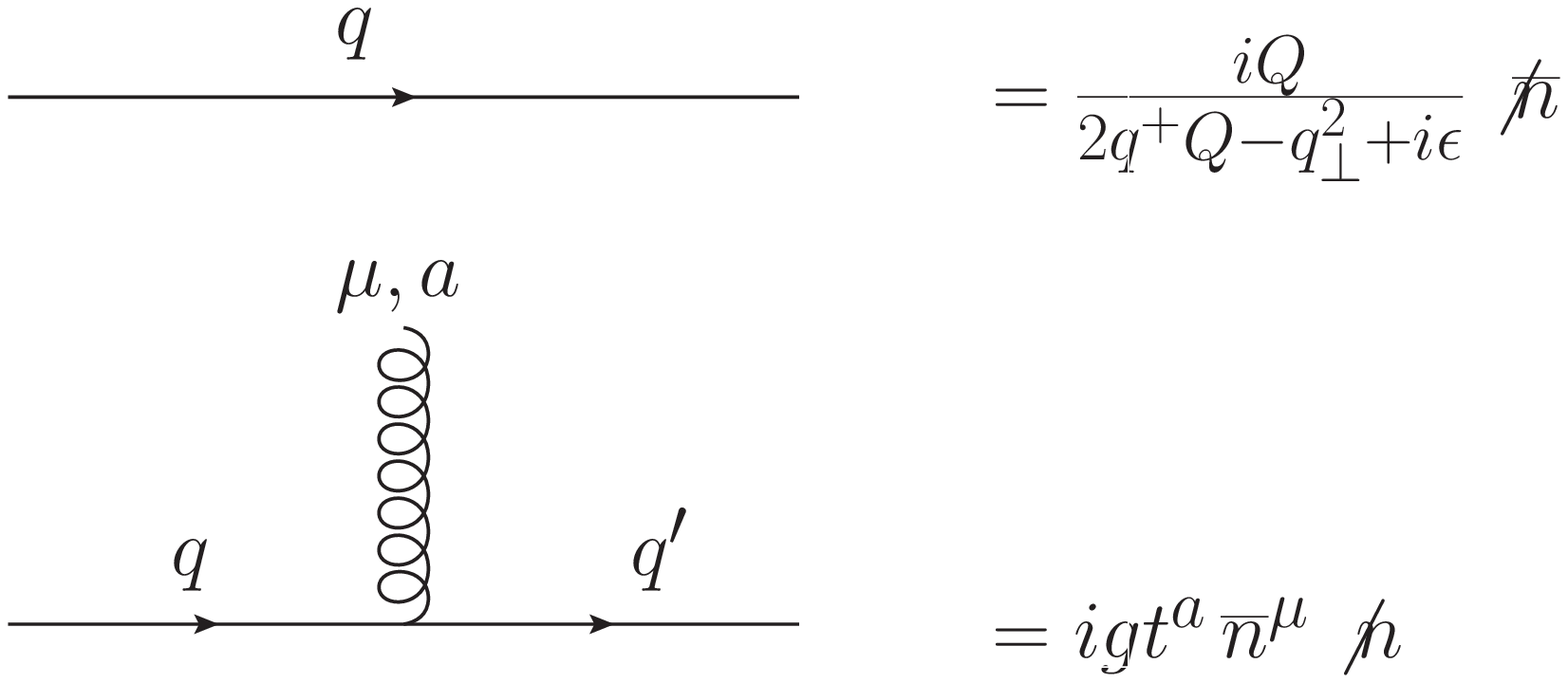}
\end{center}
\caption{Feynman rules for collinear quarks interacting with Glauber gluons.}
\label{fig:quark}
\end{figure}

\begin{figure*}[t]
 \begin{center}
\includegraphics[scale=0.5]{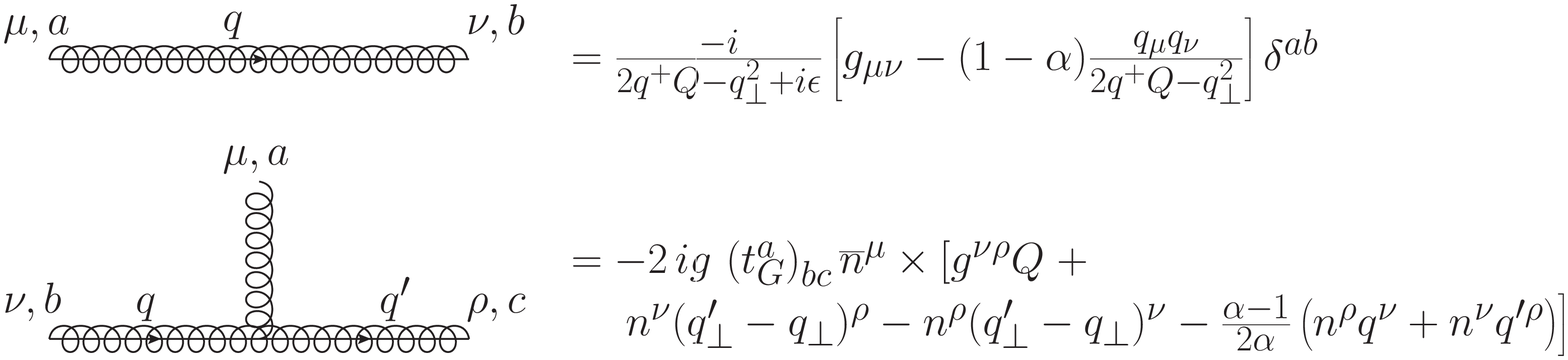}
\end{center}
\caption{Feynman rules for collinear gluons interacting with Glauber gluons in any covariant gauge. (In Feynman gauge, the gauge parameter is $\alpha = 1$.)
$t_G^{a}$ are the $SU(N)$ generators in the adjoint representation.}
\label{fig:glue}
\end{figure*}

Consider a quark that is collinear along the $-$ direction, with momentum scaling as
\begin{equation}
q = \left(q^+, q^{-}, {q}_{\perp}\right) =  Q \left(\lambda^2, 1, \lambda \right)\ .
\end{equation}
As usual in SCET, we divide the quark field $\xi(x)$ into a sum of a big component $\xi_{\bar{n}}(x)$ and a small component $\xi_{n}(x)$:
\begin{equation} \label{sepS}
\xi(x) = \xi_{\bar{n}}(x) + \xi_{n}(x), 
\end{equation}
with
\be
 \xi_{\bar{n}}(x) \equiv \frac{\slashed{\bar{n}}\slashed{n}}{2} \xi(x),\qquad
\xi_{n}(x) \equiv \frac{\slashed{n}\slashed{\bar{n}}}{2} \xi(x)\ ,\label{sepSProjection}
\ee
where we have defined the light cone vectors $n^\mu$ and $\bar n^\mu$ in (\ref{nDef}).
The small component $\xi_n (x)$ can be integrated out by using its equations of motion from
the QCD Lagrangian for the quark sector,
\begin{equation}
\mathcal{L}_{QCD} = \bar{\xi} i \slashed{D} \xi, \qquad D_\mu \equiv \p_\mu - i g A_\mu\ ,
\label{LQCD}
\end{equation}
yielding the following effective Lagrangian for $\xi_{\bar n} (x)$:
\begin{equation}
\mathcal{L}_{\bar{n}} = \bar{\xi}_{\bar{n}}\, i \slashed{n} \left(\bar{n} \cdot D\right) \xi_{\bar{n}} +
\bar{\xi}_{\bar{n}}\,i \slashed{D}_{\perp} \frac{1}{2 \, i n \cdot D} \, i \slashed{D}_{\perp} \,  \slashed{n} \, \xi_{\bar{n}} \ ,
\label{EL}
\end{equation}
where we shall only consider the case where $A_\mu$ describes a Glauber gluon, as we discussed in Section II.
Since the interaction between a collinear quark and Glauber gluons can change ${q}_\perp$ but cannot change $q^-$,
it is convenient to define 
$\xi_{\bar{n},{q}_{\perp}} (x)$ via~\cite{Bauer:2000ew,Idilbi:2008vm}
\begin{equation}
\xi_{\bar{n}} (x) = e^{- i Q x^+} \sum_{{q}_{\perp}} e^{i {q}_{\perp} \cdot {x}_{\perp}} \xi_{\bar{n}, {q}_{\perp}} (x)\ .
\label{xiexpansion}
\end{equation}
Note that derivatives acting on $\xi_{\bar{n}, {q}_{\perp}} (x)$ give only $O(\lambda^2)$ contributions and note also that
\begin{equation}
\not\!\bar{n} \, \xi_{\bar{n},{q}_{\perp}} =
 \bar \xi_{\bar{n},{q}_{\perp}} \not\!\bar{n}  = 0
\label{xiden}
\end{equation}
as a result of the projection~\eqref{sepSProjection}.
Plugging the field expansion (\ref{xiexpansion}) into the Lagrangian~(\ref{EL}) we obtain
\begin{eqnarray} \label{EL1}
\mathcal{L}_{\bar{n}} &=&
\sum_{{q}_\perp,{q}'_{\perp}}e^{i({q}_\perp-{q}'_\perp)\cdot {x}_\perp}\,
\bar{\xi}_{\bar{n},{q}_{\perp}'} \Biggl[i \left(\bar{n} \cdot D\right)\\
&+& \left(\slashed{q}_{\perp} + i \slashed{D}_{\perp}\right) \frac{1}{2 \left(Q + i n \cdot D\right)} \left(\slashed{q}_{\perp} + i \slashed{D}_{\perp}\right)\Biggr] \slashed{n} \xi_{\bar{n},{q}_{\perp}}\ .\nonumber
\end{eqnarray}
Next, we wish to expand the Lagrangian~\eqref{EL1} to lowest order in $\lambda$, but before we can do that we need to determine how the effective theory fields scale with $\lambda$. For the collinear fermion $\xi_{\bar{n}}$ we can derive its
scaling by counting powers of $\lambda$ in its propagator, as is standard in SCET~\cite{Bauer:2000ew}, and we find
\begin{equation}
\xi_{\bar{n}} \sim Q^{3/2}\lambda\ .
\end{equation}
We derive the scaling of  the gluon field $A_{\mu}$ in the Glauber region of momentum space
as in  Ref.~\cite{Idilbi:2008vm}, which is to say by thinking of $A_\mu$ as being sourced by a current consisting of the generic excitations in the medium.
In our case, working in the rest frame of the thermal medium, the generic thermal partons have momenta of order $Q(\lambda,\lambda,\lambda)$ whereas in the Breit-frame analysis of Ref.~\cite{Idilbi:2008vm} generic partons have momenta of order $Q(1,\lambda^2,\lambda)$. This means that although we can use the arguments of Ref.~\cite{Idilbi:2008vm} we get a different result. In any covariant gauge we find the scaling
\be
A_\mu \sim Q(\lambda^2, \lambda^2, \lambda^2) \  
\label{eq:GlauberScaling}
\ee
for the Glauber gluon fields. A detailed derivation is given in Appendix~\ref{app:B}.
With the scaling of the effective theory fields in hand, we can now expand the Lagrangian (\ref{EL1}) order by order in $\lambda$. We find
\be
\mathcal{L}_{\bar{n}} = \mathcal{L}_{\bar{n}}^{\mathcal{O}(\lambda^4)} + \mathcal{L}_{\bar{n}}^{\mathcal{O}(\lambda^5)} + \mathcal{O}\left(\lambda^6\right) \ ,
\ee
with the leading order term in the Lagrangian taking the form
\be
\mathcal{L}_{\bar{n}}^{\mathcal{O}(\lambda^4)} = \sum_{{q}_\perp,{q}'_{\perp}}e^{i({q}_\perp-{q}'_\perp)\cdot {x}_\perp}\,
\bar{\xi}_{\bar{n},{q}_{\perp}'} \left[i \bar{n} \cdot D + \frac{q_{\perp}^2}{2 Q} \right] \slashed{n} \xi_{\bar{n},{q}_{\perp}} \ 
\label{ELlambda4}
\ee
and the next-to-leading order term being given by
\be
\begin{split}
\mathcal{L}_{\bar{n}}^{\mathcal{O}(\lambda^5)} = & \,  
\sum_{{q}_\perp,{q}'_{\perp}} 
e^{i({q}_\perp-{q}'_\perp)\cdot {x}_\perp}\, \\ &
\frac{1}{2 Q}\bar{\xi}_{\bar{n},{q}_{\perp}'} \left[ g \left(\slashed{q}'_{\perp} \slashed{A}_{\perp} + \slashed{A}_{\perp} \slashed{q}_{\perp} \right)
+ g^2 A_\perp^2 \right] \slashed{n} \xi_{\bar{n},{q}_{\perp}} \ .\label{ELlambda5}
\end{split}
\ee
For our purposes, only the leading order Lagrangian (\ref{ELlambda4}) is needed; 
from (\ref{ELlambda4}) we obtain the Feynman rules shown in Fig.~\ref{fig:quark}. 
We have given the next-to-leading order Lagrangian (\ref{ELlambda5}) solely for the purpose
of demonstrating explicitly that
even though the perpendicular component of the Glauber field  $A_\perp$ scales with $\lambda$ in the same way that $A^+$ does, see (\ref{eq:GlauberScaling}),
 the interaction between collinear quarks and $A_\perp$ arises only at $\mathcal{O}(\lambda^5)$. 
 Because these interactions are suppressed by one power of $\lambda$ relative to the interactions in (\ref{ELlambda4}), they can safely be neglected in our analysis.

If instead we consider the case where the incident hard parton is a collinear gluon, whose components scale as
 \be
 \left(A^{+}_{\bar{n},\, q},A^{-}_{\bar{n},\, q},A^{\perp}_{\bar{n},\, q}\right)
 \sim Q\left(\lambda^2, 1, \lambda \right)\ ,
 \ee
the discussion is similar.
In any covariant gauge, we find the propagator of a collinear gluon 
 and its interaction vertex with a Glauber gluon are those given in Fig.~\ref{fig:glue}.
 The terms in the second line in the Feynman rule for the triple gluon interaction 
in Fig.~\ref{fig:glue} were absent in the first version of our paper.
They were pointed out  subsequently in Ref.~\cite{Ovanesyan:2011xy}.   None of them makes any contribution to any of our results. Neither does the $\alpha$-dependent term in the gluon propagator.

\section{Evaluating the forward scattering amplitude}

We are now ready to perform the explicit evaluation of the contribution to the forward scattering amplitude given by the diagram in Fig.~\ref{fig:forw} using the Feynman rules derived in the previous section.  The hard parton in the diagram can be a collinear quark or a collinear gluon.  We shall present the quark case explicitly, and state the result for a hard gluon at the end. This is the most technical section of our paper; the reader not interested in technicalities can safely skip to the result.

The Feynman rules derived in Section IV assume relativistic normalization of the states, whereas we need to convert to box normalization in order to obtain $P(k_\perp)$ using the optical theorem as described in Section III.  Comparing the standard relativistic normalization
\begin{equation}
\langle \alpha, {\rm relativistic} | \alpha', {\rm relativistic}\rangle = (2\pi)^3\, 2 E_\alpha \,\delta^3\left(\mathbf{q}_\alpha - \mathbf{q}_{\alpha'}\right)
\end{equation}
to the box normalization (\ref{BoxNormalization}) determines that
the box normalized states are given by
\begin{equation}
|\alpha\rangle = \frac{1}{\sqrt{2 E_\alpha L^3}}|\alpha, {\rm relativistic}\rangle \ ,
\end{equation}
meaning that the box normalized matrix elements are given by
\begin{equation}
M_{\alpha\beta}=\left.\frac{1}{2 L^3 \sqrt{E_\alpha E_\beta}}M_{\alpha\beta}\right|_{\rm relativistic}\ .
\end{equation}
We have $E_\alpha=E_\beta=Q/\sqrt{2}$, and therefore
\begin{equation}
\frac{d^2 \mathcal{A}_{m n}}{d^2 k_{\perp}} = \frac{1}{\sqrt{2} Q L^3} \left.\frac{d^2 \mathcal{A}_{m n}}{d^2 k_{\perp}}\right|_{{\rm relativistic}} \ .
\label{eq:fromreltobox}
\end{equation}

Using the Feynman rules given in Fig.~\ref{fig:quark}, the amplitude corresponding to Fig~\ref{fig:forw} can be written as
\begin{widetext}
\begin{equation}
\begin{split}
\frac{d^2 \mathcal{A}_{m n}}{d^2 k_{\perp}}= & \frac{1}{\sqrt{2} Q L^3}  \int \frac{d k^+ d k^-}{(2\pi)^2} \prod_{i=1}^{n-1} \frac{d^4 q_{i}}{(2\pi)^4} \prod_{j=1}^{m-1}  \frac{d^4 q^\prime_{j}}{(2\pi)^4}  \\
 & \times \bar{\xi}_{\bar{n}}(q_0')
  \prod_{j=m-1}^{1}
\left[
(-i g) A^{+}(-p'_{j}) \,\slashed{n}\,\frac{- i Q}{2Q q^{\prime\,+}_{j}-
q^{\prime\,2}_{j\,\perp} - i \epsilon}\,\slashed{\bar{n}}
\right]
(-i g) A^{+} (-p_{m}')\slashed{n}   \\
&  \times  2 \pi Q \delta\left(2k^{+}Q-k_{\perp}^2\right) \slashed{\bar{n}} \;
 i g A^{+} (p_{n})\slashed{n} \prod_{i=1}^{n-1} \left[\frac{ i Q}{2Qq_{i}^{+}- q_{i\,\perp}^2 + i \epsilon}\,\slashed{\bar{n}}\, i g A^{+} (p_{i})\slashed{n}\right]
  \xi_{\bar{n}}(q_0)
\end{split} \label{key1}
\end{equation}
\end{widetext}
with
\begin{equation}
p_i = q_{i} - q_{i-1} \quad {\rm for}\quad i=1 , \dots, n-1 
\end{equation}
and $p_n = k - q_{n-1}$,
\begin{equation}
p_i^{\prime} = q_{i}^{\prime} - q_{i-1}^{\prime} \quad {\rm for} \quad i=1 , \cdots, m-1 
\end{equation}
and $p_m^{\prime} = k - q_{m-1}^{\prime}$, 
\begin{equation}
\gamma^+ = \bar{n} \cdot \gamma = \slashed{\bar{n}}\,  , \qquad \gamma^- =  n \cdot \gamma =  \slashed{n}\ ,
\end{equation}
and with $A^{+}$ representing $A^{a +}\,t^a_F$, where $t^a_F$ are the $SU(N)$ generators in the fundamental representation. In the products in (\ref{key1}) the gauge fields are understood to be ordered as
\be
 \prod_{j=m-1}^{1} A^+ (-p_j') \equiv A^+(-p_{1}') \cdots A^+ (-p_{m-1}') 
 \ee
 and
 \be
 \prod_{i=1}^{n-1} A^+ (p_i) \equiv A^+(p_{n-1}) \cdots A^+ (p_1)\ .
\ee
Note additionally that in writing~\eqref{key1} we have used 
 \be
  2 \pi Q \,\th(k^+)\, \delta\left(2k^{+}Q-k_{\perp}^2\right) \slashed{\bar{n}}
  \ee
in place of a propagator at the cutting line. And, in~\eqref{key1} there are only $m+n-1$ independent momentum integrations since there is one overall constraint coming from conservation of momentum.
Finally, let us stress that 
we are treating the $A_\mu (p)$ as background fields and will average them
(say over a thermal ensemble) only at the end.  We are doing the full calculation of the propagation of the hard parton in a given background field configuration; the fact that the medium these fields describe is (or isn't) strongly coupled only becomes relevant once one does the averaging over an ensemble of field configurations.

After performing an average over the color indices, we can rewrite~\eqref{key1} as
\begin{widetext}
\begin{equation} \label{maex}
\begin{split}
\frac{d^2 \mathcal{A}_{m n}}{d^2 k_{\perp}} & = \frac{E_{mn}}{\sqrt{2}}  \int \frac{d k^+ d k^-}{(2\pi)^2} \prod_{i=1}^{n-1} \frac{d^4 q_{i}}{(2\pi)^4} \prod_{j=1}^{m-1}  \frac{d^4 q^\prime_{j}}{(2\pi)^4} \, \Tr \le[ \prod_{j=m}^{1}(-i g) A^{+}(-p'_{j})\,\prod_{i=1}^{n} i g A^{+} (p_{i})\ri]
 \\&
\times 2 \pi Q \,\delta\left(2k^{+}Q - k_{\perp}^2\right)\,\prod_{j=m-1}^{1} \left[\,\frac{- i Q}{2Qq^{\prime\,+}_{j}-
q^{\prime\,2}_{j\,\perp} - i \epsilon}\right]
\prod_{i=1}^{n-1} \left[\frac{ i Q}{2Qq_{i}^{+}- q_{i\,\perp}^2 + i \epsilon}\right]
\end{split}
\end{equation}
\end{widetext}
with
 \be
 E_{mn} \equiv \frac{1}{Q L^3\,N_c}\bar{\xi}_{\bar{n}}(q_0')\left(\slashed{n} \slashed{\bar{n}}\right)^{m-1}\,\slashed{n}\slashed{\bar{n}}\slashed{n} \left(\slashed{\bar{n}}\slashed{n}\right)^{n-1}\xi_{\bar{n}}(q_0) \ .
 \label{MmnDefn}
 \ee
Using equation~\eqref{xiden}, $\slashed{\bar{n}} \slashed{n} + \slashed{n}\slashed{\bar{n}}=2$, and equation~\eqref{spinE} from Appendix~\ref{app:A}, we find that
\begin{equation}\label{MmnResult}
E_{mn} = \frac{2^{n+m-1}}{Q L^3\,N_c} 
\;\bar{\xi}_{\bar{n}}(q_0') \,\slashed{n}\, \xi_{\bar{n}}(q_0) = \frac{1}{L^3\,N_c} 2^{n+m}\ .
\end{equation}
We have also used the fact that before the hard parton interacts with the first Glauber gluon it has momentum $q_0=Q(0,1,0)$ and the fact that for forward scattering $q_0= q_0^{\prime}$. Because $q_{0\perp}=q_{0\perp}^\prime=0$,
\begin{equation}
\sum_{i=1}^n p_{i\perp}=\sum_{i=1}^m p^\prime_{i\perp}=k_\perp\ .
\end{equation}
Note that all the $p_{i\perp}$'s and $p^\prime_{i\perp}$'s are of order $\lambda Q=T$ in magnitude, while the typical value of their sum $k_\perp$ may turn out to be larger:  since $\hat q$ is the mean $k_\perp^2$ picked up per distance travelled, the typical value of $k_\perp^2$ after a hard parton has travelled a distance $L$ is $\hat q L$, meaning that $k_\perp^2$ grows with $L$.

It is now convenient to introduce
\begin{eqnarray}
A^+ (p_i) &=& \int d^4 y_i \, e^{i p_i y_i} \, A^+ (y_i), \nonumber\\
A^+ (-p_j') &=& \int d^4 y_j' \, e^{-i p_j' y_j'} \, A^+ (y_j')\ ,\label{Eq5.11}
\end{eqnarray}
and note that
\begin{eqnarray}
& &\sum_{i=1}^n p_i y_i - \sum_{j=1}^m p_j' y_j'
= -q_0 \cdot y_1 + q_0' \cdot y_1'+ k\cdot \left(y_{n}-y_{m}'\right)  \nonumber\\
& &\quad+ \sum_{i=1}^{n-1} q_{i} \cdot \left(y_{i}-y_{i+1}\right)
- \sum_{j=1}^{m-1} q^\prime_{j} \cdot \left(y^\prime_{j}-y^\prime_{j+1}\right) \ .\label{Eq5.12}
\end{eqnarray}
Using (\ref{MmnResult}), (\ref{Eq5.11}) and (\ref{Eq5.12}), the result (\ref{maex}) becomes
\begin{equation} \label{maex2}
\begin{split}
\frac{d^2 \mathcal{A}_{m n}}{d^2 k_{\perp}}= & \frac{2^{n+m}}{\sqrt{2} L^3\,N_c} \, \int \prod_{i=1}^{n} d^4 y_{i}  \prod_{j=1}^{m} d^4 y'_{j} \,e^{-i q_0 \cdot (y_1 - y_1')} \, \\
&\times \Tr \le[\prod_{j=m}^{1}(-i g) A^{+}(y'_{j})\,\prod_{i=1}^{n} i g A^{+}(y_{i}) \ri] \\
& \times g (y_n - y_m',k_\perp) \\
&\times \prod_{j=1}^{m-1} f^* (y_j' - y_{j+1}') \prod_{i=1}^{n-1} f (y_i - y_{i+1})\ , \end{split}
\end{equation}
where we have defined
\begin{equation} \label{fDefn}
\begin{split}
f (z) & \equiv  \int {d^4 q \ov (2 \pi)^4} \, \frac{i Q}{2Q q^{+} - q_{\perp}^2 + i \epsilon} \, 
e^{i q \cdot z} \\
& = \delta(z^+) \theta(-z^-)  \frac{i Q}{4 \pi z^-} e^{- i \frac{Q}{2 z^-} z_\perp^2}, 
\end{split}
\end{equation}
and
\be
\label{gDefn}
\begin{split}
 g (z,k_\perp) & \equiv  \int {dk^+ dk^- \ov (2 \pi)^2} \, 2 \pi Q \delta\left(2k^{+}Q-k_{\perp}^2\right) \, e^{ik \cdot z}\\ 
 & = \frac{1}{2} \delta (z^+)  e^{- i k_\perp \cdot z_\perp +  i \frac{k_{\perp}^2}{2 Q} z^-} \ .
 \end{split}
\ee

So far, we have not used the $Q\rightarrow\infty$ limit anywhere in the calculation itself, although of course we used it in setting up the problem.  We shall now use the fact that both $f(z)$ 
and $g(z,k_\perp)$ simplify in this limit, once we make our statement of this limit precise.  As long as 
\begin{equation}
Q\gg p_\perp^2 z^-\ ,
\end{equation}
where $p_\perp^2\sim T^2$ is the typical magnitude of the $p_{i\perp}$'s and $p^\prime_{i\perp}$'s, the function $f(z)$ from (\ref{fDefn}) becomes
\begin{equation}
f(z) \approx \frac{1}{2} \delta\left(z^+\right) \theta\left(- z^-\right) \delta^2\left(z_{\perp}\right)\ .
\label{fsimplified}
\end{equation}
And, as long as
\begin{equation}
Q\gg k_\perp^2 z^-\ ,
\label{PreciseLimitFirstVersion}
\end{equation}
the function $g(z,k_\perp)$ from (\ref{gDefn}) becomes
\begin{equation}
g(z,k_\perp) \approx \frac{1}{2} \delta\left(z^+\right) e^{- i k_{\perp} \cdot z_{\perp}} \ .
\label{gsimplified}
\end{equation}
Since (\ref{PreciseLimitFirstVersion}) is the stronger of the two criteria, if (\ref{PreciseLimitFirstVersion}) is satisfied both $f$ and $g$ simplify.  Note that when $f(z)$ and $g(z,k_\perp)$ are employed in (\ref{maex2}) their argument $z^-$ cannot be larger than $L^-\equiv \sqrt{2} L$,
the distance along the lightcone that the hard parton travels through the medium, 
corresponding to travelling a distance $L$.
In order to guarantee that $f$ and $g$ simplify we shall therefore henceforth require that
 \be\label{PreciseLimit}
 Q \gg k_\perp^2 L \sim \hat q L^2\ .
  \ee
The physical significance of this criterion emerges upon analyzing the implications of the fact that in (\ref{fsimplified}) the function
$f(z)$ is proportional to $\delta^2(z_\perp)$. This means that in the regime (\ref{PreciseLimit})
the distance $L$ that the hard parton propagates through the medium is short enough that the trajectory of the hard parton in position space remains well-approximated as a straight line, even though it picks up transverse momentum.  We see this explicitly by
plugging the expressions (\ref{fsimplified}) and (\ref{gsimplified})  into~\eqref{maex2} and evaluating the integrals with delta functions, which set 
 \be
 y_i^+ = y_j'^+ \equiv y^+, \qquad y_{i \perp} \equiv y_\perp, \qquad y_{j \perp}'
 \equiv y_\perp'\ .
 \ee
We find
\begin{widetext}
  \bea
 {d^2 \mathcal{A}_{n m} \ov d^2 k_\perp}  &&= \frac{\sqrt{2}}{L^3\,N_c}\int dy^+  d y_\perp d y_\perp' \, e^{- i k_{\perp} \cdot (y_\perp - y'_\perp)} \,  \prod_{i=1}^{n} d y_{i}^-  \prod_{j=1}^{m}  dy_{j}'^-  \, \cr
&& \times  \; \Tr \le[ \th (y_m'^- - y_{m-1}'^-)    \cdots \th (y_2'^- - y_1'^-) \,
  (-i g) A^{+}(y^+, y_1'^-, y_{ \perp}') \cdots (-i g) A^{+}(y^+, y_m'^-, y_{ \perp}') \ri. \cr
&& \le. \times \,
\th (y_n^- - y_{n-1}^-) \cdots \th (y_2^- - y_1^-) \, i g A^{+}(y^+, y_n^-, y_{\perp})
\cdots i g A^{+}(y^+, y_1^-, y_{\perp}) \ri]\ .
\label{expor}
 \eea

Now, summing over all $m$ and $n$, we obtain
\be \label{fourP}
 \sum_{m=1,n=1}^\infty  {d^2 \mathcal{A}_{n m} \ov d^2 k_\perp}
 = \frac{\sqrt{2}}{L^3\,N_c} \int dy^+  d y_\perp d y_\perp' \, e^{- i k_{\perp} \cdot (y_\perp - y'_\perp)} \, \vev{\Tr \le[\left(W_F^\da [y^+, y_\perp']-1\right) \, \left(W_F [y^+, y_\perp]-1\right) \ri]}
\ee
 \end{widetext}
where we have introduced  the fundamental Wilson line along the
lightcone
\begin{equation}
W_F\left[y^+, y_{\perp}\right] \equiv P \left\{ \exp \left[i g \int_{0}^{L^-} d y^{-}\, A^+ (y^+, y^-,y_{\perp})\right] \right\}
\end{equation}
with $P$ denoting path-ordering, 
and where we have now restored the expectation value in the
medium. Recall that to this point we have been calculating how the hard parton propagates through one background gauge field configuration.  Now that in (\ref{fourP}) we have pushed this calculation through to the point that the gauge field appears only in the Wilson lines along the lightcone, we can complete the story by averaging over gauge field configurations.  
If the medium is quark-gluon plasma in equilibrium, then the average represented by $\langle \ldots \rangle$ is a thermal average.  In our derivation of (\ref{fourP}) it makes no difference whether the medium is strongly coupled or weakly coupled; this distinction, or indeed any properties of the medium, only become relevant when one seeks to evaluate the thermal average.

We have made a leap in going from (\ref{expor}), in which the gluon fields $A^+$ describe Glauber gluons, to (\ref{fourP}), in which we are taking a thermal average over all gluon fields.  By gauge invariance, we know that (\ref{fourP}) must be the correct generalization of (\ref{expor}) in the present context. But, in future when the effects of soft gluons and radiation (processes 1 and 3 from Section II) are computed, it is possible that additional separately gauge invariant contributions to transverse momentum broadening may arise.  As an aside, note that the expression (\ref{fourP}) is valid in any covariant gauge but not, for example, in a lightcone gauge in which $A^+=0$ and the light-like Wilson lines in (\ref{fourP}) are given by the identity.  Upon redoing the calculation in such a gauge, (\ref{fourP}) would contain the expectation value of a transverse Wilson line joining the ends of the two light-like Wilson lines.  

Because the medium is translation-invariant, the expectation value of the trace of the product of Wilson lines that arises in (\ref{fourP}) must be independent of $y^+$ and can only depend on 
the difference $y_\perp - y_\perp'$.
Upon making the change of variables
 \be
 X_\perp = \ha (y'_{\perp}+y_{\perp}) , \qquad x_\perp = y_{\perp} -y_{\perp}',
 \ee
we find
 \begin{eqnarray} 
& &\sum_{m=1,n=1}^\infty  {d^2 \mathcal{A}_{n m} \ov d^2 k_\perp} 
 = a  \int d^2 x_\perp \, e^{- i k_{\perp} \cdot x_\perp} \qquad\qquad \qquad\nonumber\\ &&\qquad \times \vev{\Tr \le[\left(W_F^\da [0,x_\perp]-1\right) \, \left(W_F [0,0]-1\right) \ri]}
\label{eq:Fint}
\end{eqnarray}
with
 \be
 a = \frac{\sqrt{2}}{L^3\,N_c} \int dy^+  d^2 X_\perp = \frac{\sqrt{2}}{L\,N_c}  \int dy^+\ .
 \ee
We now have to find a way to regularize the integral over $y^+$. We assume that we throw particles toward the medium for a time interval $\Delta t$, which is arbitrary and much larger than the box size $L$. We have normalized our states such that they describe one particle per volume $L^3$.  And,  the particles they describe move at the speed of light. Therefore, the incident flux is $1/L^3$. The total number of particles which propagate through the medium in the time interval $\Delta t$ is then given by
\begin{equation}
 \frac{1}{L^3} \, L^2\, \Delta t = \frac{\Delta t}{L}\ ,
\end{equation}
which means that in order  to obtain the probability distribution for a single particle to acquire transverse momentum $k_\perp$ we must divide (\ref{eq:Fint}) by $\Delta t/L$. The integral over $y^+$ is the projection of the time interval along the $y^+$-axis, namely $\Delta t/\sqrt{2}$, and we have
\begin{equation}
a \rightarrow \frac{L}{\Delta t} \, a = \frac{\sqrt{2}}{\Delta t \,N_c}  \int dy^+ = \frac{1}{N_c}\ .
\end{equation}
We therefore finally obtain
 \begin{eqnarray} 
&& \sum_{m=1,n=1}^\infty  {d^2 \mathcal{A}_{n m} \ov d^2 k_\perp} 
 =  \frac{1}{N_c} \int d^2 x_\perp \, e^{- i k_{\perp} \cdot x_\perp}\qquad\qquad\qquad\nonumber \\ &&\qquad\times \vev{\Tr \le[\left(W_F^\da [0,x_\perp]-1\right) \, \left(W_F [0,0]-1\right) \ri]}\ .
 \label{eq:finally}
\end{eqnarray}

Recall from (\ref{eq:amp1}) that the forward scattering amplitude which appears in the unitarity relation (\ref{eq:imp1}) is given by 
\begin{equation}
2\,\Im M_{\alpha\alpha} = \int \frac{d^2 k_\perp}{(2\pi)^2}\sum_{m=1,n=1}^\infty  {d^2 \mathcal{A}_{n m} \ov d^2 k_\perp} \ .
\label{eq:Recall}
\end{equation}
As anticipated in Section III, we can now use the unitarity relation (\ref{eq:imp1}) as well as (\ref{eq:sumbeta}), (\ref{eq:finally}) and (\ref{eq:Recall})
to identify 
 \begin{eqnarray} 
&& |M_{\beta\alpha}|^2
 =  \frac{1}{L^2\,N_c} \int d^2 x_\perp \, e^{- i k_{\perp} \cdot x_\perp} \qquad\qquad\qquad\nonumber \\ && \quad\times\vev{\Tr \le[\left(W_F^\da [0,x_\perp]-1\right) \, \left(W_F [0,0]-1\right) \ri]}.
 \label{MatrixElementSquared}
\end{eqnarray}
We now have all the ingredients in place to use (\ref{eq:PvsS}) to obtain the probability distribution $P(k_\perp)$.  
We find
\begin{equation} \label{provb1}
P(k_\perp) = \int d^2 x_{\perp} \, e^{-i k_{\perp} \cdot x_{\perp}}\,
 \sW_F (x_\perp)
\end{equation}
where we have defined
\be \label{wexp}
\sW_F [x_\perp] \equiv {1 \ov N_c} \vev{\Tr \le[ W_F^\da [0,x_\perp] \, W_F [0,0] \ri]}\ .
\ee
To demonstrate that (\ref{provb1}) is correct it suffices to check first that $P(k_\perp)/L^2$ and (\ref{MatrixElementSquared}) are identical when $k_\perp\neq 0$, which is the case since their difference is proportional to $\delta^2(k_\perp)$, and second that (\ref{provb1}) 
is correctly normalized as in (\ref{eq:Pnormalization}), which is the case since 
\begin{equation}
\int \frac{d^2 k_\perp}{(2\pi)^2}  \int d^2 x_{\perp} \, e^{-i k_{\perp} \cdot x_{\perp}}\,
 \sW_F (x_\perp)=\sW_F(0)=1.
 \end{equation}
 It is also straightforward to check that 
 \begin{equation}
 2\,\Im M_{\alpha\alpha} = 2 - \frac{1}{N_c}\left\langle \Tr W^\dagger_F[0,0] + \Tr W_F[0,0] \right\rangle
 \end{equation}
 and
  \begin{eqnarray}
\frac{ P(0)}{L^2}&=&\left|S_{\alpha\alpha}\right|^2 \nonumber\\
&=&
\left|M_{\alpha\alpha}\right|^2 + \frac{1}{N_c}\left\langle  \Tr W^\dagger_F[0,0] + \Tr W_F[0,0] \right\rangle -1
\nonumber\\
&=& 1-2\,\Im M_{\alpha\alpha}+\left|M_{\alpha\alpha}\right|^2\ ,
\end{eqnarray}
as in (\ref{eq:PvsS}).  
The expression  (\ref{provb1}) with (\ref{wexp}) is  our central technical result.

The analysis of this section can be applied completely analogously to the case in which the hard parton is a collinear gluon, instead of a collinear quark.
The only changes are that
$A^+$ is now in the adjoint representation and the $1/N_c$ factor in~\eqref{wexp}
becomes ${1 \ov N_c^2 -1}$. We conclude that 
whether the hard parton is a collinear quark or gluon, the probability distribution takes the form
\begin{equation} \label{provb}
P(k_\perp) = \int d^2 x_{\perp} \, e^{-i k_{\perp} \cdot x_{\perp}}\,
 \sW_\sR (x_\perp)
\end{equation}
with
 \be \label{Wils}
   \sW_{\sR} (x_\perp)  = \frac{1}{d\left(\mathcal{R}\right)} 
   \vev{\Tr \le[ W_{\sR}^\da [0,x_\perp] \, W_{\sR} [0,0] \ri]}
  \ee
where $\mathcal{R}$ is the $SU(N)$ representation to which the collinear particle belongs 
and $d\left(\mathcal{R}\right)$ is  the dimension of this representation.  
Eq.~\eqref{provb} is an elegant expression saying that the probability for the 
quark to obtain transverse momentum $k_\perp$ is simply given by the Fourier transform in $x_\perp$ of the expectation value~\eqref{Wils} of two light-like Wilson lines separated in the transverse plane by the vector $x_\perp$.
Eq.~\eqref{provb} has been obtained previously 
by Casalderrey-Solana and Salgado
and by Liang, Wang and Zhou 
using different methods~\cite{CasalderreySolana:2007zz,Liang:2008vz}.

\section{$\hat q$ from light-like Wilson lines}

The jet quenching parameter $\hat q$ is the mean transverse momentum picked up by the hard parton per unit distance travelled, or equivalently per unit time.  We reproduce its definition (\ref{qhatFirstTime}) here:
\begin{equation} \label{qhatD}
\hat{q} \equiv \frac{1}{L} \int \frac{d^2 k_{\perp}}{(2\pi)^2}\,k_{\perp}^2 \, P(k_{\perp})\ .
\end{equation}
Substituting our result (\ref{provb}) for $P(k_\perp)$ in (\ref{qhatD}), we find that
\begin{equation}
\hat{q} = \frac{\sqrt{2}}{L^-}  \, \int \frac{d^2 k_{\perp}}{(2\pi)^2}\,k_{\perp}^2 \, \int d^2 x_{\perp} \, e^{-i k_{\perp} \cdot x_{\perp}}\,
\sW_\sR (x_\perp) \ ,
\label{qhatFromW}
\end{equation}
where we have replaced $L$ by the distance along the lightcone $L^-=\sqrt{2}L$.   Upon evaluating $\hat q$ in Section VII, we shall see that it is $L^-$-independent.

The probability distribution (\ref{provb}), derived in a covariant gauge, contains only the two Wilson lines along the light cone. Our field theory set-up and SCET calculation require $L^- T \gg 1$, and we can therefore consider the two light-like  Wilson lines to have infinite length. 
In this limit, 
it does not matter how we join the two light-like Wilson lines at infinity 
along the transverse direction in order to make the gauge invariance of our result manifest, because in a covariant gauge the contribution of these short transverse  segments is subleading.  
If we were to make a gauge transformation from covariant gauge to a lightcone gauge in which $A^+=0$, all of the contributions of the medium to the Wilson loop would then be encoded in the two short transverse Wilson line segments.  In such a gauge, these short transverse Wilson line segments would make a contribution to the (logarithm of) the expectation value of the Wilson loop that is proportional to the extent $L^-$ of the medium.
This indicates that repeating our calculation in lightcone gauge would be quite inconvenient, and is the reason we see no motivation for doing so.

\begin{figure}
 \begin{center}
\includegraphics[scale=0.55]{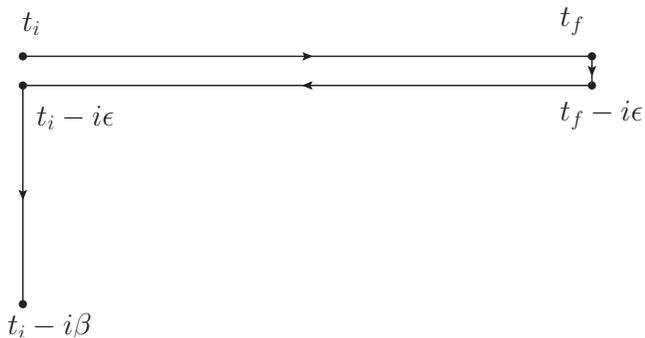}
\end{center}
\caption{Schwinger-Keldysh contour that must be used in the evaluation of $\sW_{\sR}(x_\perp)$.}
\label{fig:skc}
\end{figure}
It is important to notice that the expectation value of the trace of the product of two light-like Wilson lines
that arises in $P(k_\perp)$ and hence in $\hat q$, namely $\sW_{\sR}(x_\perp)$ of (\ref{Wils}), has a different operator ordering from that in a standard Wilson loop.  Each of the $A^+$'s in (\ref{expor}) can be written as the product of an operator and a group matrix: $A^+=(A^+)^a t^a$.  It is clear from the explicit expression (\ref{expor}) that in $\sW_{\sR}(x_\perp)$ both the operators and the group matrices are path ordered.  In contrast, in a conventional Wilson loop the group matrices are path ordered but the operators are time ordered.  Because the operators in (\ref{Wils}) are path ordered, the expectation value in (\ref{Wils}) should be described using the 
 Schwinger-Keldysh contour in Fig.~\ref{fig:skc} with one of the light-like Wilson lines on the ${\rm Im}\, t=0$ segment of the contour and the other light-like Wilson line on the ${\rm Im}\, t=-i\epsilon$ segment of the contour.  The infinitesimal displacement of one Wilson line with respect to the other in Fig.~\ref{fig:skc} ensures that the operators from the two lines are ordered such that all operators from one line come before any operators from the other.
In contrast,  the loop $\sC$ for a standard Wilson loop operator
lies entirely at ${\rm Im} \, t =0$, and the operators for a standard Wilson loop are time ordered.   This distinction was not made in the early work that related 
$\hat q$ to a light-like Wilson loop~\cite{Wiedemann:2000za,Kovner:2003zj} or in the more 
recent analyses of Refs.~\cite{Liu:2006ug,Liu:2006he} in which this Wilson loop (with conventionally time ordered operators) was evaluated in the plasma 
of strongly coupled ${\cal N}=4$ SYM theory.   Seeing now that the property of the medium which controls $\hat q$ is in fact $\sW_{\sR}(x_\perp)$, with its path ordered operators, we must
 revisit the calculation of $\hat q$ in strongly coupled ${\cal N}=4$ SYM theory. We shall do so in Section VII, finding that the calculation changes but the result does not.

Before turning to the strong coupling calculation of $\sW_{\sR}(x_\perp)$ and hence $\hat q$, it is worth casting the relation between them in a form which yields further intuition.
It has been argued
in Refs.~\cite{Kovner:2003zj,Liu:2006ug,Liu:2006he} that in the adjoint representation
 \be \label{qhatE} 
 \sW (x_\perp) = \exp\left[- {1 \ov 4 \sqrt{2}}\, \hat{q}\,L^{-}\,x^2_{\perp} \right]\ .
 \ee
We can use our result (\ref{provb}) to check that this expression is self-consistent.
If we substitute (\ref{qhatE}) into~\eqref{provb} we obtain
 \begin{equation} \label{probD}
P (k_\perp) = \frac{4 \sqrt{2} \pi}{\hat{q}\, L^{-}} \, \exp\left[-\frac{\sqrt{2} k^2_{\perp}}{\,\hat{q}\,L^-}\right]\ ,
\end{equation} 
which we can then substitute into 
the definition of $\hat q$, \eqref{qhatD}, which becomes $\hat q=\hat q$, confirming that
the expression~\eqref{qhatE} is indeed self-consistent. 
Note that the expression~\eqref{probD} has a simple physical interpretation: the probability that the quark has gained 
transverse momentum  $k_\perp$ is given by  diffusion in momentum space 
with a diffusion constant $D$ given by
 \be
 D = \hat q L\ .
 \ee
 This is indeed consistent with the general physical expectation stated earlier: 
 the effect of small kicks due to Glauber gluons is that the quark performs 
 Brownian motion in momentum space even though in coordinate space it remains 
 on a light-like trajectory.

\section{Computation of $\hat q$ in strongly coupled $\sN=4$ Supersymmetric Yang Mills theory
revisited}

We now compute~\eqref{Wils} for $\sN=4 $ SYM theory 
with gauge group $SU(N_c)$ in the large $N_c$
and strong coupling limit using its gravitational dual~\cite{AdS/CFT}, namely 
the AdS Schwarzschild black hole at nonzero temperature. 

\begin{figure}
 \begin{center}
\includegraphics[scale=0.50]{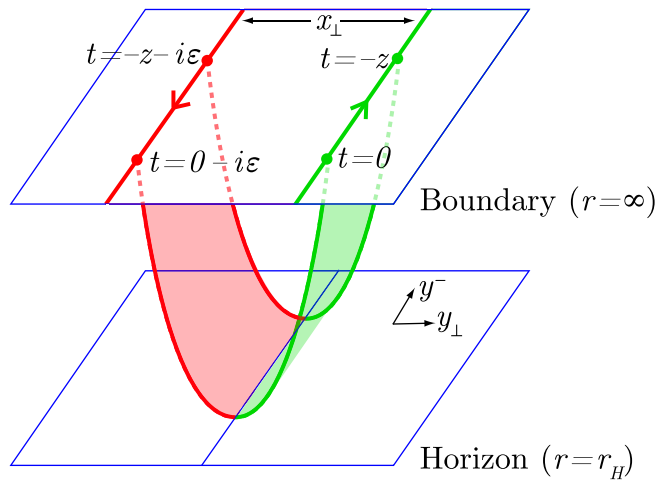}
\end{center}
\caption{String configuration for the thermal expectation value of~\eqref{Wils}.}
\label{fig:horizon}
\end{figure}

Let us first recall the standard AdS/CFT procedure for computing a Wilson loop 
in the large $N_c$ and strong coupling limit~\cite{Rey:1998ik}, which was applied to a light-like Wilson loop (with standard operator ordering; not that in~\eqref{Wils}) in Refs.~\cite{Liu:2006ug,Liu:2006he}. Consider a Wilson loop operator $W(\sC)$ specified by a closed loop $\sC$ in the $(3+1)$-dimensional field theory, and thus on the boundary of the $(4+1)$- dimensional AdS space.
$\langle W(\sC)\rangle$  is then given by the exponential of the classical action of an extremized string worldsheet $\Sig$ in AdS which ends on $\sC$. The contour ${\cal C}$ lives within the $(3+1)$-dimensional
Minkowski space boundary, but the string world sheet $\Sig$ attached to it hangs ``down'' into the bulk of the curved five-dimensional AdS$_5$ spacetime.
More explicitly, consider a Wilson loop made from two long parallel light-like Wilson lines separated by a distance $x_\perp$ in a transverse direction.  (The string world sheet hanging down into the bulk from these two Wilson lines can be visualized as in Fig.~\ref{fig:horizon} below if one keeps everything in that figure at ${\rm Im} \, t =0$, i.e.~if one ignores the subtlety that is the reason we are revisiting this calculation.)
Upon parameterizing the two-dimensional world sheet by the coordinates $\sigma^{\alpha}=(\tau,\sigma)$,
the location of the string world sheet in the five-dimensional spacetime with coordinates
$x^M$ is
\begin{equation}
    x^M = x^M(\tau,\sigma)\, 
    \label{para}
\end{equation}
and the Nambu-Goto action for the string world sheet is given
by
 \begin{equation}
S =- \frac{1 }{ 2 \pi \alpha'} \int d\sigma d \tau \, \sqrt{ - \det
g_{\alpha \beta}}\, . \label{ngac}
 \end{equation}
Here,
\begin{equation}
  g_{\alpha \beta} = G_{MN} \partial_\alpha x^M \partial_\beta x^N\,
  \label{inm}
\end{equation}
is the induced metric on the world sheet and $G_{MN}$ is the metric of the
$(4+1)$-dimensional AdS$_5$ spacetime. Denoting by $S (\sC)$ the classical action which  extremizes the Nambu-Goto action~\eqref{ngac} for the string worldsheet with the boundary condition that it ends on the curve ${\cal C}$, 
the expectation value of the Wilson loop operator is then given by
 \begin{equation}
\langle{W({\cal C})}\rangle = \exp\left[ i\, \lbrace S ({\cal C})  - S_0 \rbrace \right] \, ,\label{exewi}
 \end{equation}
where the subtraction
$S_0$ is the action of two disjoint strings hanging straight down from the two Wilson lines. In order to obtain the thermal expectation value at nonzero temperature, one takes 
the metric in~\eqref{inm} to be that of an AdS Schwarzschild black hole,
 \begin{eqnarray}
ds^2 & = &  G_{MN} dx^M dx^N  \nonumber\\
&=& - f dt^2 + \frac{r^2 }{ R^2} (dx_1^2 + dx_2^2 + dx_3^2)+
\frac{1 }{ f} dr^2 \, ,
\label{metr1}
\end{eqnarray}
where $R$ is the curvature radius of the AdS space and where we have defined
\begin{equation}
f \equiv \frac{r^2 }{ R^2} \left(1 - \frac{r_H^4 }{ r^4} \right)\, .
\label{feff}
 \end{equation}
Here, $r$ is the  coordinate of the 5th dimension and the black hole
horizon is located at $r = r_H$. The temperature $T$ of the Yang-Mills theory plasma is given by the Hawking temperature of the black hole, $T=r_H/(\pi R^2)$, and $R$ and the string tension $1/(2\pi\alpha')$ are related to the 't Hooft coupling\footnote{As is conventional in the two relevant bodies of literature, we shall use $\lambda$ to denote both the 't Hooft coupling and the SCET expansion parameter.  The context will make clear which we mean where.}
in the Yang-Mills theory $\lambda\equiv g^2 N_c$ by $\sqrt{\lambda} = R^2/\alpha'$.

 \begin{figure*}[t]
 \begin{center}
\includegraphics[scale=0.55]{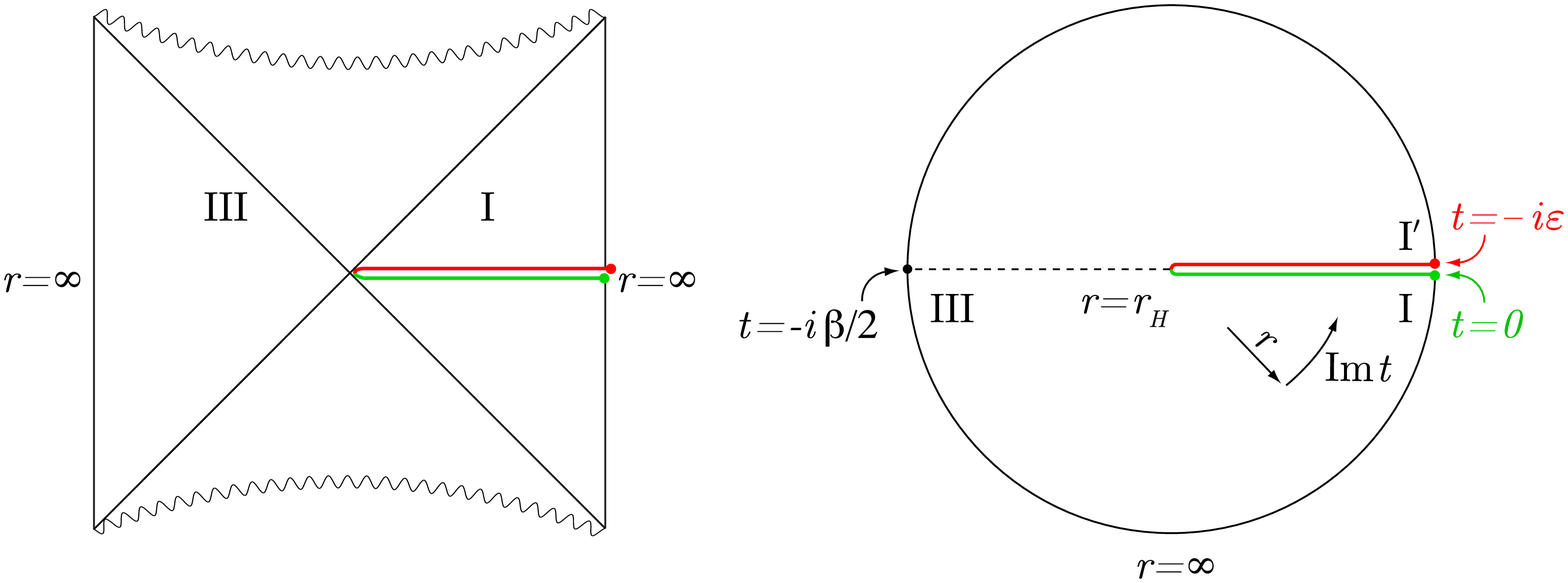}
\end{center}
\caption{Penrose diagrams for Lorentzian (${\rm Im}\,t=0$) and Euclidean (${\rm Re}\,t=0$) sections of an AdS black hole. In the right panel, the two light-like Wilson lines are points at $r=\infty$, indicated by colored dots.  These dots are the boundaries of a string world sheet that extends inward to $r=r_H$, which is at the origin of the Euclidean section of the black hole. In the left panel, the string world sheet and its endpoints at $r=\infty$ are shown at ${\rm Re}\, t=0$;  as ${\rm Re}\,t$ runs from $-\infty$ to $\infty$, the string world sheet sweeps out the whole of quadrant $I$.  }
\label{fig:disk}
\end{figure*}

The thermal expectation value of a light-like Wilson loop at strong coupling was computed as we have just described in Refs.~\cite{Liu:2006ug,Liu:2006he}. We shall quote the result below (after demonstrating that it is unchanged.) We comment here, however, that in Ref.~\cite{Liu:2006ug} it was found by calculation that the extremized string world sheet, hanging from the light-like Wilson lines at the boundary, extends all the way down to the black hole horizon.  We shall see below that in the computation of (\ref{Wils}) this is mandatory: it can be inferred without calculation that the extremized string world sheet must reach the black hole horizon.

We now consider the computation of~\eqref{Wils}, with its nonstandard operator ordering corresponding, as we have discussed in Section VI, to putting one of the two light-like Wilson lines on the ${\rm Im} \, t =0$ contour in Fig.~\ref{fig:skc} and the other on the ${\rm Im}\, t=-i\epsilon$ contour.  The procedure we shall describe is a specific example of the
more general discussion of Lorentzian
AdS/CFT given recently by Skenderis and Van Rees in Refs.~\cite{Skenderis:2008dh,Skenderis:2008dg,vanRees:2009rw}, which we have followed.  However, we shall describe the construction in a self-contained fashion that (we hope) will make it sound obvious to any reader familiar with the evaluation of standard Wilson loops using AdS/CFT.

In order to compute~\eqref{Wils} we first need to construct the bulk geometry corresponding to the ${\rm Im}\,t=-i\epsilon$ segment of the Schwinger-Keldysh contour 
in Fig.~\ref{fig:skc}. For this purpose it is natural to consider the black hole geometry with complex time, i.e. treating the time coordinate $t$ in the metric~\eqref{metr1} as a complex variable
In Fig.~\ref{fig:disk}, we show two slices of this complexified geometry. The left plot is the Penrose diagram for the fully extended black hole spacetime with quadrant $I$ and $III$ corresponding to the slice ${\rm Im} \, t =0$ and $\Im t = -{\beta \ov 2}$ respectively, while the right plot is for the Euclidean black hole geometry, i.e. corresponding to the slice $\Re t =0$. Note that because the black hole has a nonzero temperature, the imaginary part of $t$ is periodic with the period given by the inverse temperature $\beta$. 
In the left plot the imaginary time direction can be considered as a circular direction coming out of the paper at quadrant $I$, going a half circle to reach quadrant $III$ and then 
going into the paper for a half circle to end back at $I$. In the right plot the real time direction can be visualized as the direction perpendicular to the paper.

The first segment of the Schwinger-Keldysh contour in Fig.~\ref{fig:skc}, with ${\rm Im}\,t=0$, lies at the boundary ($r=\infty$) of quadrant $I$ in Fig.~\ref{fig:disk}. 
The second segment of the Schwinger-Keldysh contour, 
with ${\rm Im}\, t=-i\epsilon$, lies at the $r=\infty$ boundary of a copy of $I$ that in the left plot of Fig.~\ref{fig:disk} lies infinitesimally outside the paper and in the right plot of Fig.~\ref{fig:disk} lies at an infinitesimally different angle.  We shall denote this copy of $I$ by $I'$. 
The geometry and metric in $I'$ are identical to those of $I$. Note that $I'$ and $I$ are joined together at the horizon $r=r_H$, namely at the origin in the right plot of Fig.~\ref{fig:disk}. Now, the thermal expectation value~\eqref{Wils} can be computed by putting the two parallel light-like Wilson lines at the boundaries of $I$ and $I'$, and finding the extremized string world sheet which ends on both of them.
Note that since $I$ and $I'$ 
meet only at the horizon, the only way for there to be a nontrivial (i.e.~connected) string world sheet whose boundary is the two Wilson lines in (\ref{Wils}) is for such a string world sheet to touch the horizon.
Happily, this is precisely the feature of the string world sheet found by explicit calculation 
in Refs.~\cite{Liu:2006ug,Liu:2006he}.  So, we can use that string world sheet in the present analysis, with the only difference being that half the string world sheet now lies on $I$ and half on $I'$, as illustrated in Fig.~\ref{fig:horizon}. 

We conclude that the result for the expectation value~\eqref{Wils}, with its nonstandard path ordering of operators, is identical to that obtained in Refs.~\cite{Liu:2006ug,Liu:2006he} for a light-like Wilson loop with standard time ordering of operators.  That is, in strongly coupled ${\cal N}=4$ SYM theory $\sW(x_\perp)$ in the adjoint representation is given by 
\begin{equation}
\begin{split}
& \mathcal{W}_{\mathcal{A}}(x_\perp) = \\
& \exp \left[ - \sqrt{2} a \sqrt{\lambda}\,  L^- T \left(\frac{\pi x_\perp T}{2 a} \sqrt{1 + \frac{4 a^2}{\pi^2 T^2 x_\perp^2}} - 1\right)\right],
\end{split}
\label{fullW}
\end{equation}
where 
\begin{equation}
a =\frac{ \sqrt{\pi}\, \Gamma(5/4) }{\Gamma(3/4)} \simeq 1.311 \,.
\end{equation}
The $x_\perp$-independent term in the exponent in (\ref{fullW}), namely ``the -1'', is the finite subtraction of $S_0$, which was identified in Ref.~\cite{Liu:2006ug} as the action of two disjoint strings hanging straight down from the two Wilson lines to the horizon of the AdS black hole.
Our calculation serves as a check of the value of $S_0$, since  
only with the correct $S_0$ do we obtain $\mathcal{W}_{\mathcal{A}}(0)=1$  and a correctly normalized probability distribution $P(k_\perp)$. 
Note that
our field theory set-up and SCET calculation require $L^- T \gg 1$, and our supergravity calculation requires $\lambda \gg 1$, meaning that our result (\ref{fullW})  is only valid for
\begin{equation}
\sqrt{\lambda} \, L^- T \gg 1 .
\label{StrongCouplingGaussian}
\end{equation}
In this regime, (\ref{fullW}) is very small unless $\pi x_\perp T/(2a)$ is small.  This means that when we take the Fourier transform of \eqref{fullW} to obtain the probability 
distribution $P(k_\perp)$, in the regime (\ref{StrongCouplingGaussian})
where the calculation is valid 
the Fourier transform is dominated by small values of $x_\perp$, for which
\begin{equation}
\mathcal{W}_{\mathcal{A}}(x_\perp) \simeq 
\exp \left[ - \frac{\pi^2} {4 \sqrt{2} a} \sqrt{\lambda} L^- T^3 x_\perp^2 \right]\ ,
\label{StrongCouplingGaussian2}
\end{equation}
and we therefore obtain
\begin{equation}
P(k_\perp) = \frac{4\sqrt{2}a}{\pi \sqrt{\lambda} T^3 L^-}\exp\left[ - \frac{\sqrt{2} a k_\perp^2}{\pi^2 \sqrt{\lambda} T^3 L^-} \right] \ .
\end{equation}
Thus, the probability distribution $P(k_\perp)$ is a Gaussian  and the jet-quenching parameter can be read off by comparing  Eqs.~(\ref{qhatE}) and (\ref{StrongCouplingGaussian2}), yielding~\cite{Liu:2006ug}
\begin{equation}
\hat q=\frac{\pi^{3/2}\Gamma(\frac{3}{4})}{\Gamma(\frac{5}{4})}\sqrt{\lambda} T^3 \ .
\label{qhatResult}
\end{equation}
This turns out to be in the same ballpark as the values of $\hat q$ inferred from RHIC data on the suppression of high momentum partons in heavy ion collisions~\cite{Liu:2006ug,Liu:2006he}.

The jet quenching parameter can then be calculated in any conformal theory with a gravity dual, yielding results just as in Ref.~\cite{Liu:2006he}.  In a large class of such theories~\cite{Liu:2006he},
\begin{equation}
\frac{\hat q_{\rm CFT}}{\hat q_{{\cal N}=4}}=
\sqrt{\frac{s_{\rm CFT}}{s_{{\cal N}=4}}}\ ,
\label{qhatversusentropy}
\end{equation}
with $s$ the entropy density.  This result makes a central qualitative lesson from (\ref{qhatResult}) clear: in a strongly coupled plasma, the jet quenching parameter is not proportional to the entropy density or to some number density of distinct scatterers.  This qualitative lesson is more robust than any attempt to make a quantitative comparison to QCD. But, we note that
if QCD were conformal, (\ref{qhatversusentropy}) would suggest 
\begin{equation}
\frac{\hat q_{\rm QCD}}{\hat q_{{\cal N}=4}} \approx 0.63\ .
\end{equation}
And, analysis of how $\hat q$ changes in a particular toy model in which nonconformality can be introduced by hand then suggests that introducing the degree of nonconformality seen in QCD thermodynamics may increase $\hat q$ by a few tens of percent~\cite{Liu:2008tz}.

Our new calculation of $\hat q$ in ${\cal N}=4$ SYM theory via~\eqref{Wils} also nicely resolves a technical subtlety in Refs.~\cite{Liu:2006ug,Liu:2006he}. It was observed there (and subsequently discussed at length in Refs.~\cite{Argyres:2006yz}) that in addition to the extremized string configuration which touches the horizon, the string action also has another trivial solution 
which lies solely at the boundary, at $r=\infty$.   Based on the connection between position in the $r$ dimension in the gravitational theory and energy scale in the quantum field theory, the authors of Ref.~\cite{Liu:2006ug,Liu:2006he} argued that physical considerations (namely the fact that $\hat q$ should reflect thermal physics at energy scales of order $T$) require selecting the extremized string configuration that touches the horizon.  Although this physical argument remains valid, we now see that it is not necessary. In (\ref{Wils}), the two Wilson lines are at the boundaries of $I$ and $I'$, with different values of ${\rm Im}\,t$.  That means that there are no string world sheets that connect the two Wilson lines without touching the horizon.  So, once we have understood how the nonstandard operator ordering in (\ref{Wils}) modifies the boundary conditions for the string world sheet, we see that the trivial world sheet of Refs.~\cite{Liu:2006ug,Liu:2006he} and all of its generalizations in Refs.~\cite{Argyres:2006yz} do not satisfy the correct boundary conditions.  The nontrivial world sheet illustrated in Fig.~\ref{fig:horizon}, which is sensitive to thermal 
physics~\cite{Liu:2006ug,Liu:2006he,Liu:2008tz}, is the only extremized world sheet bounded by the two light-like Wilson lines in (\ref{Wils}).  The result (\ref{qhatResult}) follows.

\section{Future Directions}

We have computed the probability density $P(k_\perp)$ that describes the transverse momentum broadening 
experienced by a hard parton moving through the quark-gluon plasma of QCD upon neglecting the possibility of gluon radiation.   Our result is (\ref{provb}), which relates $P(k_\perp)$ to the expectation value of two light-like Wilson lines $\sW(x_\perp)$ in (\ref{Wils}).  In turn, the jet quenching parameter $\hat q$ can be obtained from $\sW(x_\perp)$ according to (\ref{qhatFromW}).  

In Section VII we have computed $\sW(x_\perp)$ in the plasma of large-$N_c$ strongly coupled ${\cal N}=4$ SYM theory.  The fact that the Wilson lines in (\ref{Wils}) have operators (and not just color matrices) that are path ordered introduces subtleties into this calculation that had not been previously taken into account.  However, these subtleties turn out not to change the results (\ref{qhatE}) for $\sW(x_\perp)$ and (\ref{qhatResult}) for $\hat q$.  In fact, the calculation has become more straightforward than previously realized, since there is only one extremized string world sheet bounded by these Wilson lines.

These results open a variety of future directions:
\begin{itemize}
\item
$\sW(x_\perp)$, and from it $\hat q$, can be computed for the plasma of QCD at high enough temperatures that physics at scales $\sim T$ is weakly coupled using the hard thermal loop effective theory.  It would be interesting to compare the value of $\hat q$ so obtained to other weak-coupling determinations of this quantity (see e.g. \cite{CaronHuot:2008ni}).
\item
Our soft collinear effective theory analysis should be extended to include the effects of radiation, allowing the analysis of parton energy loss in addition to transverse momentum broadening.  When a collinear quark radiates a collinear gluon, all three collinear particles (incoming quark, outgoing quark, radiated gluon) will experience transverse momentum broadening due to interactions with Glauber gluons that we have calculated.  It is therefore reasonable to expect that $\hat q$ will enter into the calculation of the spectrum of the radiated gluons and thus the parton energy loss, as is the case in other formalisms in which these quantities are calculated. It will be very interesting to see how $\hat q$ enters into the soft collinear effective theory analysis of parton energy loss.
\item
Our soft collinear effective theory analysis should be extended to include the interaction of the hard parton with soft gluons from the plasma, in addition to the Glauber gluons that we have focussed on.  These effects are suppressed by a power of a perturbative $\alpha$ and so are subleading in the limit of infinite parton energy, but because there are more soft gluons in the plasma than Glauber gluons these effects (and radiation) should be included in the analysis before any comparison with data is attempted.
\item
The full power of soft collinear effective theory lies in its systematic organization of higher order corrections.  This potential motivates the whole approach and so of course should be investigated, but only after the directions above have been pursued.
\item
We have calculated $P(k_\perp)$ in QCD and determined that it is related to 
$\sW(x_\perp)$; we have then calculated $\sW(x_\perp)$ 
in strongly coupled ${\cal N}=4$ SYM theory.   
It would be interesting to use gauge/gravity duality to 
compute $P(k_\perp)$ itself for a quark moving through the ${\cal N}=4$ SYM plasma with $v\rightarrow 1$, to see that it is given by (\ref{provb}).  
\item
The calculation of $\sW(x_\perp)$ that we have done for a quark with any finite mass $M$ 
moving with $v\rightarrow 1$
can easily be extended to the case of Wilson lines that are not quite light-like, with a velocity $v$ that is less than 1 but that satisfies
$(1-v^2)^{1/4} < \sqrt{\lambda}T/M$~\cite{Liu:2006he}. 
But, our formulation of $P(k_\perp)$ in terms of $\sW(x_\perp)$ relies on the $Q\rightarrow\infty$ limit and so does not extend to this regime.
Some progress has been made in the opposite kinematic regime, namely
for a quark moving through the plasma with a velocity $v$ that can be as small as zero and is not too large: $(1-v^2)^{1/4} > \sqrt{\lambda}T/M$.  This regime incorporates the limit in which one first takes $M\rightarrow\infty$ and only then allows $v\rightarrow 1$.
Although $P(k_\perp)$ has also not been calculated in this regime,  
the analogue of $\hat q$, which is conventionally denoted $\kappa_T$, 
has been computed in ${\cal N}=4$ SYM theory in Refs.~\cite{CasalderreySolana:2006rq,Gubser:2006nz,CasalderreySolana:2007qw}.   ($\kappa_T$ describes the mean transverse momentum picked up per distance travelled for quarks in this kinematic regime, but it has not been related to  jet quenching.)
The formalism used to compute $\kappa_T$ in this different kinematic regime is quite different from ours. It starts from the physics of a heavy quark diffusing in {\it position} space, whereas in the limit in which we do our calculation the quark follows a straight line trajectory in position space, diffusing only in transverse momentum.  This different physical picture yields a different calculation, phrased in terms of correlation functions of local operators inserted along a single Wilson line.  Not surprisingly, the 
result for $\kappa_T$ is quite different from (\ref{qhatResult}).  Clearly, the $v\rightarrow 1$ and $M\rightarrow \infty$ limits do not commute. And yet, Casalderrey-Solana and Teaney derive a field theoretical expression for $\kappa_T$ that is very similar to our (\ref{qhatFromW})~\cite{CasalderreySolana:2007qw}.  Perhaps this is a hint that a common formalism for the calculation of $P(k_\perp)$ at all $v$ can be found. This remains an open problem.
\end{itemize}

\acknowledgments

We acknowledge helpful conversations with Jorge Casalderrey-Solana, Bob Jaffe, Ahmad Idilbi, Chris Lee, Abhijit Majumder, Juan Maldacena, Claudio Marcantonini, Grigory Ovanesyan, Massimiliano Procura, Dam Son, Iain Stewart, Balt van Rees, X. Wang and Urs A. Wiedemann. The work of HL was supported in part by a DOE Outstanding Junior Investigator grant.
This research was
supported in part by the DOE Offices of Nuclear and High Energy
Physics under grants \#DE-FG02-94ER40818 and  \#DE-FG02-05ER41360.

\appendix

\section{The fermion bilinear $\bar{\xi}_{s,\bar{n}}(q) \gamma^- \xi_{s,\bar{n}}(q)$}
\label{app:A}

In this Appendix we evaluate the fermion bilinear $\bar{\xi}_{s,\bar{n}}(q) \gamma^- \xi_{s,\bar{n}}(q)$ that arises in Section V. We use the Dirac basis for the gamma matrices
\begin{equation}
\gamma^0 = \left(\begin{array}{cc}
\mathbf{1} & 0 \\
0 & - \mathbf{1} \\
\end{array}
\right) \qquad \gamma^i = \left(\begin{array}{cc}
0 & \sigma^i \\
-\sigma^i & 0 \\
\end{array}
\right)
\end{equation}
with 
\begin{equation}
\gamma^{+} =  \slashed{\bar{n}} = \frac{\gamma^0 + \gamma^3}{\sqrt{2}} = \frac{1}{\sqrt{2}}\left(\begin{array}{cc}
\mathbf{1} & \sigma^3 \\
-\sigma^3 & -\mathbf{1}  \\
\end{array}
\right)
\end{equation}
and
\begin{equation}
\gamma^{-} =  \slashed{n} = \frac{\gamma^0 - \gamma^3}{\sqrt{2}} = \frac{1}{\sqrt{2}}\left(\begin{array}{cc}
\mathbf{1} & -\sigma^3 \\
\sigma^3 & -\mathbf{1}  \\
\end{array}
\right) \ .
\label{gamma+-}
\end{equation}

It is straightforward to find the spinor wave function for a collinear quark with four momentum 
\begin{equation}
q = \left(0,Q,0\right)
\end{equation}
with $Q \gg m$  by solving the free Dirac equation, obtaining
\begin{equation}
 \xi_{s,\bar{n}}(q) = \sqrt{q^0}
\left(
\begin{array}{c}
\chi_{s} \\
- \sigma^3 \chi_{s}
\end{array}
\right)\ ,
\label{xinbar}
\end{equation}
where $\chi_s$ is a two-component spinor normalized as 
\begin{equation}
 \chi_{r}^{\dag}\chi_{s} = \delta_{rs} \ .
\end{equation}

Now, using (\ref{xinbar}) and (\ref{gamma+-}) and suppressing the spin indices, we find
\begin{eqnarray}
&&\bar{\xi}_{s,\bar{n}}(q)\gamma^- \xi_{s,\bar{n}}(q)\nonumber\\ 
&&= \frac{q^0}{\sqrt{2}} \left(
\begin{array}{cc}
\chi^{\dag}_{s} & - \sigma^3 \chi^{\dag}_{s}
\end{array}
\right) \left(\begin{array}{cc}
\mathbf{1} & -\sigma^3 \\
-\sigma^3 & \mathbf{1}  \\
\end{array}
\right)
\left(
\begin{array}{c}
\chi_{s} \\
- \sigma^3 \chi_{s}
\end{array}
\right)\nonumber\\
 &&= 2 \sqrt{2} q^0\ ,
\end{eqnarray}
namely
\begin{equation} \label{spinE}
 \bar{\xi}_{s,\bar{n}}(q) \gamma^- \xi_{s,\bar{n}}(q) = 2 q^- \ ,
\end{equation}
which is the result we use in going from (\ref{MmnDefn}) to (\ref{MmnResult}).

\section{Scaling of the Glauber field}
\label{app:B}

In this Appendix, we derive the scaling of the Glauber gluon field $A_\mu$ with $\lambda$. Following the method described in  Ref.~\cite{Idilbi:2008vm}, we derive the scaling of $A_{\mu}$ by analyzing the linear response formula
\be
A_{\mu}^a(x) = \int d^4 y \, G_{\mu\nu}^{a b} (x-y) \, J^{b\,\nu}(y) \ \label{eq:LinearResponse}
\ee
which describes the Glauber gluon field sourced by the current $J^{b\,\nu}$ that is composed of generic excitations of the medium. Here, 
$G_{\mu\nu}^{a b}$ is the Glauber gluon propagator,
which is given at tree-level in any covariant gauge by
\be
G_{\mu\nu}^{a b} (x-y) = \delta^{ab} \, \int \frac{d^4 p}{(2\pi)^4} \, \frac{- i g_{\mu\nu}}{p^2 + i \epsilon} \, e^{- i p \cdot (x-y)} \ .
\ee
For the purpose of determining the scaling of the Glauber gluon field 
$A_\mu$, it suffices to consider the 
contribution to the current $J^{b\,\nu}$ coming from the quarks $\psi$
in the medium, namely
\be
J^{b\,\nu}(y) = \bar{\psi}(y) \gamma^{\nu} t^b \psi(y) \ .
\ee
We expand the field $\psi$ into its Fourier components
\be
\begin{split}
\psi(y) = & \, \int \frac{d^3 q}{(2 \pi)^3 \sqrt{2 E_{\mathbf{q}}}}  \\ &
\sum_{s=1,2} \left[b^s_{\mathbf{q}} u^s(q) e^{-i q \cdot y} + 
d^{s\,\dag}_{\mathbf{q}} v^s(q) e^{i q \cdot y}\right]_{q^0= E_{\mathbf{q}}} \ ,
\end{split}
\ee
where $E_{\mathbf{q}} = \left|\mathbf{q} \right|$, and the domain for the integration over $d^3 q$ is $\mathbf{q} \sim \lambda Q$, since we are in the medium frame. Thus, the 
expression (\ref{eq:LinearResponse}) for the Glauber gluon field includes the term
\be
\begin{split}
A_{\mu}^a(x) = & \, \int \frac{d^3 q \, d^3 p}{(2\pi)^6 \,\sqrt{2 E_{\mathbf{q}}}\,\sqrt{2 E_{\mathbf{q} - \mathbf{p}}}} \, 
  \frac{- i \, e^{- i p \cdot x}}{p^2 + i \epsilon} \\ &
\sum_{r,s} b^{r\,\dag}_{\mathbf{q} - \mathbf{p}} b^s_{\mathbf{q}} \, \bar{u}^r(q - p) \, \gamma_\mu t^a u^s(q) \ ,
\end{split}
\label{eq:Aexpr}
\ee
as well as other terms that scale the same way.
Here, $\mathbf{p}$ has support in the Glauber region.

We can now use Eq.~(\ref{eq:Aexpr}) to determine how $A_\mu$ scales with $\lambda$. The power counting of the $b^s_{\mathbf{q}}$ operator is obtained from the anti-commutation relation
\be
\left\{b^r_{\mathbf{q}}, b^s_{\mathbf{q}^\prime} \right\} = \delta^{rs} (2\pi)^3 \delta^3(\mathbf{q} - \mathbf{q}^\prime) \ , 
\ee
which implies that $b^r_{\mathbf{q}} \sim \lambda^{-3/2} Q^{-3/2}$. Likewise, the scaling of the spinor $u^s(q)$ can be obtained from the completeness relation
\be
\sum_{s=1,2} u^s(q)\bar{u}^s(q) = \slashed{q} \ ,
\ee
which implies that $u^s(q)\sim \lambda^{1/2} Q^{1/2}$.
Knowing the scaling of 
$b^s_{\mathbf{q}}$ and $u^s(q)$,  and 
using $d^3 q \sim \lambda^3 Q^3$ and $E_\mathbf{q}\sim E_{\mathbf{q}-\mathbf{p}}\sim \lambda Q$ as appropriate for generic modes from the medium and $d^3 p\sim \lambda^4 Q^3$ and $p^2\sim \lambda^2 Q^2$ as appropriate for Glauber modes,
the scaling of each of the components of the Glauber gluon field $A_\mu$ now follows from
Eq.~(\ref{eq:Aexpr}). We obtain
\be
A_{\mu} \sim Q \left(\lambda^2,\lambda^2,\lambda^2\right) \ ,
\ee
namely Eq.~\ref{eq:GlauberScaling}.

\end{document}